\documentclass[12pt]{iopart}
\pdfoutput=1
\usepackage[pdftex]{graphics}
\usepackage{color}

\usepackage{graphicx} 
\usepackage{relsize}

\def\beqra{\begin{eqnarray}}
\def\eeqra{\end{eqnarray}}
\def\beq{\begin{equation}}
\def\eeq{\end{equation}}
\def\ds{\displaystyle}

\def\bk{{\bf k}}
\def\vp{\varphi}

\def\bx{{\bf{x}}}
\def\bp{{\bf{p}}}
\def\bq{{\bf{q}}}
\def\re#1{(\ref{#1})}

\def\half{\mbox{\small$\frac{1}{2}$}}

\def\alt{\stackrel{<}{\sim}}
\begin{document}

\title{Next-to-leading resummations in cosmological perturbation theory}
\author{Stefano Anselmi}
\address{Dipartimento di Fisica ``G. Galilei'', Universit\`a di Padova, via Marzolo 8, I-35131, Padova, Italy\\
INFN, Sezione di Padova, via Marzolo 8, I-35131, Padova, Italy}
\ead{stefano.anselmi@pd.infn.it}
\author{Sabino Matarrese}
\address{Dipartimento di Fisica ``G. Galilei'', Universit\`a di Padova, via Marzolo 8, I-35131, Padova, Italy\\
INFN, Sezione di Padova, via Marzolo 8, I-35131, Padova, Italy}
\ead{sabino.matarrese@pd.infn.it}
\author{Massimo Pietroni}
\address{INFN, Sezione di Padova, via Marzolo 8, I-35131, Padova, Italy}
\ead{massimo.pietroni@pd.infn.it}

\begin{abstract}
One of the nicest results in cosmological perturbation theory is the analytical resummaton of the leading corrections at large momentum, which was obtained by Crocce and Scoccimarro for the propagator in \cite{Crocce:2005xz}. Using an {\it exact} evolution equation, we generalize this result, by showing that a class of next-to-leading corrections can also be resummed at all orders in perturbation theory. The new corrections modify the propagator by a few percent in the Baryonic Acoustic Oscillation range of scales, and therefore cannot be neglected in resummation schemes aiming at an accuracy compatible with future generation galaxy surveys. Similar tools can be employed to derive improved approximations for the Power Spectrum.
\end{abstract}

\maketitle
\section{Introduction}
Cosmological perturbation theory (hereafter, PT, for a review, see \cite{Bernardeau:2001qr}) has attracted a renewed interest in the recent literature. On the one hand, future galaxy surveys will measure -- at the percent level-- the statistical properties of matter distribution in a range of scales and redshifts in which linear perturbation theory (LT) predictions for the power spectrum (PS) and higher order correlation functions are not more accurate than O(10\%). On the other hand, due to practical limitations in machine-time, N-body simulations are far from being the ideal tool to scan over parameters and models, and involve larger complexity when extra components (such as massive neutrinos, a non-minimally coupled quintessence field, ...) or non-gaussian initial conditions are considered. Moreover, a series of works, pioneered by  Ref.~\cite{Crocce:2005xz} by Crocce and Scoccimarro (hereafter, CS), has successfully investigated the possibility of improving PT  by ``resumming" perturbative contributions at all orders. Different resummation schemes have been proposed and their results for the PS in the $\Lambda$CDM cosmology have been compared to N-body simulations in the Baryon Acoustic Oscillations range of scales \cite{Valageas:2006bi, McDonald:2006hf, MP07a, Matarrese:2007wc,Crocce:2007dt, Matsubara:2007wj, Pietroni:2008jx, Matsubara:2008wx, Nishimichi:2008ry,Taruya:2009ir,Hiramatsu:2009ki,Taruya:2010mx,Valageas:2010yw}, showing in most cases an agreement in the few percent range. These methods have been applied also to less standard --although interesting -- cosmological scenarios, such as mixed dark matter ({\it i.e.} cold dark matter + massive neutrinos)  \cite{Lesgourgues:2009am}, coupled dark matter-dark energy models \cite{Saracco:2009df}, and models with various types of primordial non-Gaussianity \cite{Bartolo:2009rb}.

The main reason for the success of these resummation methods, compared to traditional PT, was nicely discussed in CS. In PT, each successive order in the expansion in powers of the linear PS becomes more and more important at small scales, or large Fourier momenta, $k$. Therefore, the perturbative expansion unavoidably breaks down at high $k$'s and low redshifts. On the other hand, the leading contributions at large $k$'s  have a simple form which, in some cases, allows them to be summed to all orders, giving a convergent result. This was shown analytically for the {\it propagator}, {\it i.e.} the -- properly normalized --  cross-correlator between the final density, or velocity, perturbations and the initial ones (for its exact meaning, see Eq.~(\ref{PSGNG})). In this case, the large-momentum leading order contribution, at $n$-th order in PT, grows as $(-1)^n k^{2n}$, but the sum exhibits a nice Gaussian damping, $\propto \exp (-k^2 \sigma^2 D^2/2)$, with $D$ the linear growing mode, and $\sigma^2$ the one-dimensional dispersion velocity (see Eq.~(\ref{sig2})). 

In other words, the bad UV behavior of PT can be greatly alleviated if one reorganizes the series expansion, for instance using as a zero order approximation the resummed propagator of CS instead of its linear approximation. This idea forms the basis of the `renormalized perturbation theory' (RPT) approach \cite{Crocce:2005xy}. Similar analytical results hold also for a special class of higher order correlation functions, the `multi-point propagators considered in \cite{Bernardeau:2008fa, Bernardeau:2010md}. On the other hand, for the most directly observable quantities such as the PS, the bispectrum, and higher order correlators, the resummation program cannot be carried out analytically, and the semi-analytical methods are needed \cite{MP07a, Matarrese:2007wc, Matsubara:2007wj,Crocce:2007dt, Pietroni:2008jx, Matsubara:2008wx, Nishimichi:2008ry,Taruya:2009ir,Hiramatsu:2009ki,Taruya:2010mx,Valageas:2010yw}.

In this paper, we will discuss how to go beyond the CS result for the propagator by taking into account subleading corrections which were neglected in the analytic resummation described above. We will identify a class of next-to-leading contributions and we will show that the resummation performed in CS can be generalized to include this larger class of perturbative corrections. Our results indicate that the new resummed propagator corrects the CS one by a few percent in the BAO range, and therefore the new contributions should be taken into account. 

The basic tool to go beyond the CS resummation is provided by {\it exact} evolution equations for the propagator, which we will derive. We will discuss in which approximation the solution of this equation gives the CS propagator and then explore two different schemes to incorporate next-to leading corrections. 

This paper is organized as follows. In Section \ref{EOM} we review the derivation of PT in the useful compact form introduced in  \cite{Crocce:2005xy}. In Section \ref{ExactEq} we introduce the generating functional for the statistical correlators, along the lines of Ref.~\cite{Matarrese:2007wc}, recall the diagrammatic language useful to discuss PT and its resummations,  and derive the exact evolution equation for the propagator. In Section \ref{FACTORIZATIONSECTION} we clarify the relation between the evolution equation and the resummation of `chain-diagrams' performed in CS, and then, in Section  \ref{FACTORIZATIONSECTIONPS} we discuss how to go beyond the CS result by taking into account the renormalized PS in the chain-diagrams. In Section \ref{results} we present our numerical results and, finally, in Section \ref{CONCL} we discuss them and give our conclusions.

\section{Nonlinear fluid equations and the Propagator}
\label{EOM}
We can write the three fluid equations (continuity, Euler, Poisson) in Einstein-de Sitter cosmology as follows
\beqra
&&\frac{\partial\,\delta_m}{\partial\,\tau}+
{\bf \nabla}\cdot\left[(1+\delta_m) {\bf v} \right]=0\,,\nonumber\\
&& \frac{\partial\,{\bf v}}{\partial\,\tau}+{\cal H}\,{\bf v} \, + ( {\bf v} 
\cdot {\bf \nabla})  {\bf v}= - {\bf \nabla} \phi\,,\nonumber\\
&&\nabla^2 \phi = \frac{3}{2}\,\,{\cal H}^2  \, \, \delta_m\, ,
\label{Euler}
\eeqra
where  ${\cal H}= d \log a/d \tau$ is the Hubble Parameter in conformal time,  while $\delta_m({\bf x},\,\tau)$ and  ${\bf v} ({\bf x},\,\tau)$ are the DM number-density fluctuation and the DM peculiar velocity field, respectively.

Defining, as usual, the velocity divergence $\theta(\bx,\,\tau) = 
\nabla \cdot {\bf v} (\bx, \,\tau)$, and going to
Fourier space, the equations in (\ref{Euler}) give
\beqra
&&\frac{\partial\,\delta_m({\bf k}, \tau)}{\partial\,\tau}+\theta({\bf k}, 
\tau) \nonumber\\
&& \qquad + \int d^3\bq\, d^3\bp \,\delta_D({\bf k}-\bq-\bp)
 \alpha(\bq,\bp)\theta(\bq, \tau)\delta_m(\bp, \tau)=0\,,\nonumber\\
&&\frac{\partial\,\theta({\bf k}, \tau)}{\partial\,\tau}+
{\cal H}\,\theta({\bf k}, \tau)  +\frac{3}{2} {\cal H}^2 \,
\delta_m({\bf k}, \tau)\nonumber\\
&& \qquad   +\int d^3\bq \,d^3\bp \,\delta_D({\bf k}-\bq-\bp) 
\beta(\bq,\bp)\theta(\bq, \tau)\theta(\bp, \tau) = 0 \,.\label{EulerFourier}
\eeqra
The nonlinearity and non-locality of the fluid equation are encoded in 
the two functions
\beq\alpha(\bq,\bp )= \frac{(\bp + \bq) \cdot \bq}{q^2}\,,\quad \quad
\beta(\bq,\bp ) = \frac{(\bp + \bq)^2 \,
\bp \cdot \bq}{2 \,p ^2 q^2}\,,
\eeq
which couple different modes of density and velocity fluctuations. 

One can write Eqs.~(\ref{EulerFourier}) in a compact form \cite{Crocce:2005xz}. First, 
we introduce the doublet $\vp_a$ ($a=1,2$), given by
\beq\left(\begin{array}{c}
\varphi_1 ( {\bf k}, \eta)\\
\varphi_2 ( {\bf k}, \eta)  
\end{array}\right)
\equiv 
e^{-\eta} \left( \begin{array}{c}
\delta_m  ( {\bf k}, \eta) \\
-\theta  ( {\bf k}, \eta)/{\cal H}
\end{array}
\right)\,,
\label{doppietto}
\eeq
where the time variable has been replaced by the logarithm of 
the scale factor,
\[ 
\eta= \log\frac{a}{a_{in}}\,,
\]
$a_{in}$ being the scale factor at a conveniently remote epoch, 
in which all the relevant scales are well inside the linear regime. 

Then, we define a {\it vertex} function, 
$\gamma_{abc}({\bf k},{\bf p},{\bf q}) $ ($a,b,c,=1,2$) 
whose only independent, non-vanishing, elements are
\beqra
&&\gamma_{121}({\bf k},\,{\bf p},\,{\bf q}) = 
\frac{1}{2} \,\delta_D ({\bf k}+{\bf p}+{\bf q})\, 
\alpha(\bp,\bq)\,,\nonumber\\
&&\gamma_{222}({\bf k},\,{\bf p},\,{\bf q}) = 
\delta_D ({\bf k}+{\bf p}+{\bf q})\, \beta(\bp,\bq)\,,
\label{vertice}
\eeqra
and 
$\gamma_{121}({\bf k},\,{\bf p},\,{\bf q})  = 
\gamma_{112}({\bf k},\,{\bf q},\,{\bf p}) $.

The two equations (\ref{EulerFourier}) can now be rewritten in a compact form as
\beq
\partial_\eta\,\varphi_a({\bf k}, \eta)= -\Omega_{ab}\,
\varphi_b({\bf k}, \eta) + e^\eta 
\gamma_{abc}({\bf k},\,-{\bf p},\,-{\bf q})  
\varphi_b({\bf p}, \eta )\,\varphi_c({\bf q}, \eta ),
\label{compact}
\eeq
where 
\beq 
\Omega= \left(\begin{array}{rr}
\ds 1 & \ds -1\\&\\
\ds -\frac{3}{2} & \ds \frac{3}{2} \end{array}
\right)\,.
\label{bigomega}
\eeq
Repeated indices are summed over, and integration over momenta $\bq$ and $\bp$ is understood.\\

To extend the validity of this approach to $\Lambda$CDM, we will reinterpret the variable $\eta$ as the logarithm of the linear growth factor of the growing mode, i.e. ~\cite{Bernardeau:2001qr,Crocce:2005xz,Pietroni:2008jx},
\beq
\eta=\ln (D/D_{in})\,,
\eeq and we redefine the field in Eq.~(\ref{doppietto}) as 
\beq 
\left(\begin{array}{c}
\varphi_1 ( {\bf k}, \eta)\\
\varphi_2 ( {\bf k}, \eta)  
\end{array}\right)
\equiv 
e^{-\eta} \left( \begin{array}{c}
\delta_m  ( {\bf k}, \eta) \\
-\theta  ( {\bf k}, \eta)/{\cal H }f
\end{array}\right)\,,
\label{PhiF}
\eeq
with $f=d \ln D/d\ln a$. As discussed in ~\cite{Pietroni:2008jx}, the above approximation is accurate at better than $1\%$ level in the whole range of redshifts and scales we are interested in. 

If we consider the linear equations (obtained in the $e^{\eta}\gamma_{abc}\rightarrow 0$ limit) we can define the {\it linear retarded propagator}  as the operator giving the evolution of the field $\varphi_{a}$ from $\eta_{in}$ to $\eta$,
\beq
\vp^{L}_{a}(\bk,\eta)=g_{ab}(\eta,\eta_{in})\vp^{L}_{b}(\bk,\eta_{in})\,.
\eeq
The linear propagator obeys the equation 
\beq
(\delta_{ab}\partial_{\eta}+\Omega_{ab})g_{bc}(\eta,\eta_{in})=\delta_{ac}\delta_D(\eta-\eta_{in}).
\eeq
with causal boundary conditions. It is given explicitly by the following expression \cite{Crocce:2005xz},
\beq 
g_{ab}(\eta_a,\eta_b) =\left[ {\bf B} + {\bf A}\, e^{-5/2 
(\eta_a -\eta_b)}\right]_{ab}\, \theta(\eta_a-\eta_b)\,,
\label{proplin}
\eeq
 with $\theta$ the 
step-function, and
\beq {\bf B} = \frac{1}{5}\left(\begin{array}{cc}
3 & 2\\
3 & 2
\end{array}\right)\,\qquad {\mathrm{and}} \qquad
{\bf A} = \frac{1}{5}\left(\begin{array}{rr}
2 & -2\\
-3 & 3
\end{array}\right)\,.
\eeq
The growing ($\vp_a \propto \mathrm{const.}$) and the decaying 
($\vp _a\propto \exp(-5/2 \eta_a)$) modes can be selected by 
considering initial fields $\vp_a$ proportional to 
\beq u_a = \left(\begin{array}{c} 1\\ 
1\end{array}\right)\,\qquad\mathrm{and} 
\qquad v_a=\left(\begin{array}{c} 1\\ -3/2\end{array}\right)\,,
\label{ic}
\eeq
respectively.

\section{The non-linear propagator and its evolution}
\label{ExactEq}

Following the path-integral formulation of cosmological perturbation theory introduced in \cite{Matarrese:2007wc} we can derive exact evolution equations for the propagator and for the PS. In this section we will give a short review of the formalism introduced and discussed in  \cite{Matarrese:2007wc}, and we will obtain the exact evolution equation for the propagator, which will be solved in various approximations in the remaining sections.

The generating functional for the time-dependent correlators between perturbations is given by
 \beqra
\label{genf}
Z[J_a,\, K_b;\,P^0] 
&=&\int {\cal D}  \vp_a {\cal D} \chi_b
\exp \biggl\{-\half \int d\eta_a d\eta_b   \chi_a P^0_{ab} 
\delta(\eta_a) \delta(\eta_b)\chi_b \nonumber\\
&+& 
i \int d\eta \left[ \chi_a g^{-1}_{ab} \vp_b  -  
e^\eta\,\gamma_{abc} \chi_a \vp_b \vp_c + 
J_a \vp_a +  K_b \chi_b\right]\biggr\} \,, 
\label{GENFUN}
\eeqra
 where $J_{a}$ and $K_{b}$ are sources for $\varphi_{a}$ and $\chi_{b}$ respectively, and $P^{0}_{ab}(k)$ is the PS at the initial time $\eta_{in}=0$. In deriving the above expression we have assumed Gaussian initial conditions. Non-Gaussian initial conditions can be taken into account by including a non-vanishing bispectrum, trispectrum, etc., in the first line of Eq.~(\ref{GENFUN}).
Derivatives of Eq.~(\ref{GENFUN}) w.r.t. the sources $J_a$ and $K_b$ give all the possible statistical correlators involving the fields $\vp_a$ and $\chi_b$. 
As usual, it is more convenient to discuss {\em connected correlators}, which can be derived from the  generating functional,
 \beq W=-i \log Z\,.
\label{WW}
\eeq
In the following, we will be interested in the PS 
\beq
\langle \vp_a(\bk,\eta_a)\vp_b(\bk^\prime,\eta_b) \rangle
 \equiv \delta_D(\bk +\bk^\prime) P_{ab}(k; \eta_a,\eta_b)\,,
 \label{PSd}
\eeq
and in the propagator,
\beq
\langle \vp_a(\bk,\eta_a)\chi_b(\bk^\prime,\eta_b) \rangle
 \equiv i\, \delta_D(\bk +\bk^\prime) G_{ab}(k;\eta_a,\eta_b)\,.
 \label{propd}
\eeq
They are given by the second derivatives of $W$, according to the relations,
\beqra
&&\left. \frac{\delta^2 W}{\delta J_a\, \delta J_b}\right|_{J_a,\,K_b=0} 
= i\, \delta_D({\bk}+{\bk}') P_{ab}\,,\nonumber\\
&&\left. \frac{\delta^2 W}{\delta J_a\, \delta K_b}\right|_{J_a,\,K_b=0} 
= - \delta_D({\bk}+{\bk}') G_{ab},\nonumber\\
&&\left. \frac{\delta^2 W}{\delta K_a\, \delta J_b}\right|_{J_a,\,K_b=0} 
= - \delta_D({\bk}+{\bk}') G_{ba} \,,\nonumber\\
&&\left. \frac{\delta^2 W}{\delta K_a\, \delta K_b}\right|_{J_a,\,K_b=0} =0 \,.
\label{w2}
\eeqra

One can also consider the {\em effective action}, built in terms of the {\em average} fields  
 \beq
\bar{\vp}_a[J_c, K_d](\bk,\eta)  \equiv \frac{\delta W[J_c, K_d]}{\delta J_a}\,,\;\;\;\;\;
\bar{\chi}_a[J_c, K_d](\bk,\eta) =\frac{\delta W[J_c, K_d]}{\delta K_b}\,,
\label{fields}
\eeq
where the functional derivatives are evaluated at generic ({\it i.e.} non-vanishing) values for the sources $J$ and $K$.
The effective action is given by the Legendre transform of $W$,
\beq
\Gamma[\bar{\vp}_a,\, \bar{\chi}_b] =  W[J_a, K_b] - \int d\eta\,d^3 \bk \left(J_a \bar{\vp}_a 
+ K_b \bar{\chi}_b \right)\,,
\label{legendre}
\eeq
and its derivatives with respect to $\bar{\vp}_{a}$ and $\bar{\chi}_{a}$ give rise to the one-particle irreducible Green functions (1PI). The two-point 1P1 functions are given by
\beqra
&& \left.\frac{\delta^2 \Gamma[\bar{\vp}_a,\bar{\chi}_b]}{\delta \bar{\vp}_a \delta \bar{\vp}_b}\right|_{\bar{\vp}_a,\bar{\chi}_b=0}=0 \,,\nonumber\\
&& \left.\frac{\delta^2 \Gamma[\bar{\vp}_a,\bar{\chi}_b]}{\delta \bar{\chi}_a \delta \bar{\vp}_b}\right|_{\bar{\vp}_a,\bar{\chi}_b=0} \equiv \left( g^{-1}_{ab}-\Sigma_{ab}\,\right)\delta_D(\bk_a+\bk_b)\, ,\nonumber\\
&& \left.\frac{\delta^2 \Gamma[\bar{\vp}_a,\bar{\chi}_b]}{\delta \bar{\vp}_a \delta \bar{\chi}_b}\right|_{\bar{\vp}_a,\bar{\chi}_b=0}\equiv \left( g^{-1}_{ba}-\Sigma_{ba}\,\right)\delta_D(\bk_a+\bk_b)\,
,\nonumber\\
&&\left.\frac{\delta^2 \Gamma[\bar{\vp}_a,\bar{\chi}_b]}{\delta \bar{\chi}_a \delta \bar{\chi}_b}\right|_{\bar{\vp}_a,\bar{\chi}_b=0}\equiv \left( i  P^0_{ab}(k) \delta(\eta) \delta(\eta_b)
 + i \Phi_{ab}\,\right)\delta_D(\bk_a+\bk_b)\,,
\label{GAMMA2}
\eeqra
where we have isolated the `free' parts, $g^{-1}_{ab}$ and $P^0_{ab}$, from the `interacting' ones, $\Sigma_{ab}$ and $\Phi_{ab}$.

The 1PI two-point functions in Eq.~(\ref{GAMMA2}) are related to connected ones in Eq.~(\ref{w2}) by the following relations, 

\beqra
P_{ab}(k; \eta_a,\eta_b) &=& G_{ac}(k;\eta_a,0)
G_{bd}(k;\eta_b,0) P^0_{cd}(k)\,\nonumber\\
&&+
 \int_0^{\eta_a} d s_1
\int_0^{\eta_b} d s_2\,
G_{ac}(k;\eta_a,s_1)
G_{bd}(k;\eta_b, s_2) 
\Phi_{cd}(k; s_1, s_2)\,,\nonumber\\&&
\label{fullP}
\eeqra
and
\beq
 G_{ab} (k; \eta_a, \eta_b) = \left[g^{-1}_{ba} - 
\Sigma_{ba}\right]^{-1}(k; \eta_a, \eta_b)\,,
\label{fullG}
\eeq
where the last expression has to be interpreted in a formal sense, that is,
\beqra
 &&G_{ab} (k;\, \eta_a, \eta_b)=g_{ab} (\eta_a, \eta_b) \nonumber\\
 &&\qquad\qquad+ \int_{\eta_b}^{\eta_a} d s_1 \int_{\eta_b}^{s_1}d s_2\, g_{ac} (\eta_a, s_1) \Sigma_{cd}(k;\, s_1, s_2) g_{db} (s_2, \eta_b) + \cdots\,.
\label{GEXPANSION}
\eeqra

While the physical meaning of the PS -- namely the cross-correlator of the density-velocity fields $\vp_a$ and $\vp_b$ computed at times $\eta_a$ and $\eta_b$ -- is clear from its definition, Eq.~(\ref{PSd}), the presence of the auxiliary field $\chi$ in the definition (\ref{propd}) makes the interpretation of the propagator more obscure. Insight on its physical meaning can be obtained from Eq.~(\ref{fullP}), by sending one of the two times, e.g. $\eta_b$ to the initial time, {\it i.e.} $\eta_b\to 0^+$,
\beq
P_{ab}(k;\eta_a,0) = G_{ac}(k;\eta_a,0) P^0_{cb}(k)+\int_{0}^{\eta_{a}}\, ds\,G_{ac}(k;\,\eta_{a},s)\,{\Phi}_{cb}(k;\,s,0)\,,
\label{PSGNG}
\eeq
where we have used the property of the full propagator $G_{bd}(k;\eta,0)\to \delta_{bd}$ for $\eta\to 0^+$, see Eqs.~(\ref{proplin}) and (\ref{GEXPANSION}). 
The second term at the RHS of the equation above vanishes if the statistics is Gaussian at the initial time, therefore, in this case, the propagator 
connects the initial PS, $P^0_{ab}(k) \equiv P_{ab}(k;\eta_a=\eta_b=0)$, to the cross-correlator between the initial fields and the `final' ones evaluated at $\eta_a >0$. 
Taking the initial PS on the linear growing mode, 
\beq
P^0_{ab}(k) \simeq P^0(k)u_a u_b,
\label{crossc}
\eeq
where the $u_a$ vector has been defined in Eq.~(\ref{ic}), we see that the late-time cross correlators are entirely given in terms of $P^0(k)$ and the two combinations,
\beq
G_a(k;\,\eta,0)\equiv G_{ac}(k;\,\eta,0) u_c,\;\;(a=1,2)\,,
\label{Gcont}
\eeq
which will be considered in the following. 

We notice here that Eq.~(\ref{PSGNG}) holds in general, {\it i.e.}, even when the initial time is taken to correspond to a lower redshift, where non-linearities and non-gaussianities have to be taken into account. We will make use of this relation in sect.~\ref{FACTORIZATIONSECTIONPS}.

\begin{figure}
\centerline{\includegraphics[width = 11cm,keepaspectratio=true]{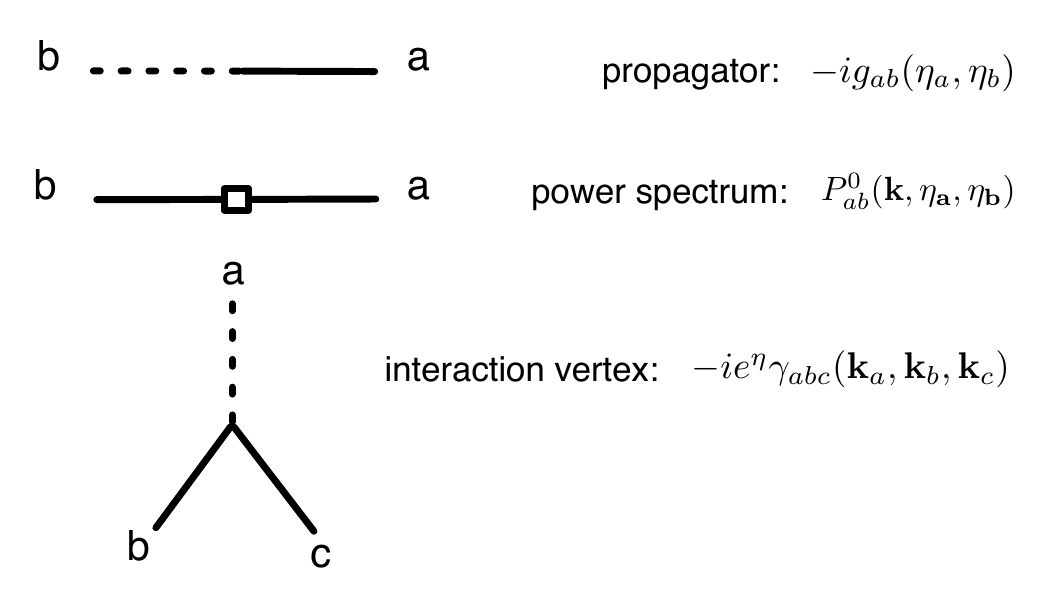}}
\caption{The Feynman Rules for cosmological perturbation theory}
\label{FEYNMAN}
\end{figure}

In general, all kind of N-point correlators, and in particular the two-point functions, can be computed in perturbation theory (PT).  From the path integral formulation discussed above one can derive Feynman rules for the building blocks of PT, namely, the free propagator, the linear PS, and the interaction vertex \cite{Matarrese:2007wc}, from which all the higher order correlators can be built. These are summarized in Fig.~\ref{FEYNMAN}: continuous and dashed lines indicate $\varphi_a$ and $\chi_a$ fields, respectively.

Equations (\ref{fullP}) and (\ref{fullG}) by themselves, however, do not rely on PT, and therefore offer the opportunity of computing the two-point correlators non-perturbatively. A convenient way to deal with the propagator in a non-perturbative way is to cast Eq.~(\ref{GEXPANSION}) in a closed form \cite{Crocce:2005xy} \footnote{In turbulence theory this result is well known, see for instance Ref.~\cite{1961AnPhy..14..143W,Procaccia:1995}.}
\beqra
 &&G_{ab}( k;\,\eta_a,\eta_b)=g_{ab}(\eta_a,\eta_b)\nonumber\\
&&\qquad\qquad\qquad+\int_{\eta_b}^{\eta_a} d s_1 \int_{\eta_b}^{s_1}d s_2\, g_{ac}(\eta_a,s_1)\Sigma_{cd}( k;\,\,s_1,\,s_2)\,G_{db}(k;\,s_2,\eta_b)\, , \nonumber\\
&&\nonumber\\
\label{FORMALGEQ}
\eeqra
which is equivalent to  \re{GEXPANSION}, as can be shown by expanding the full propagator $G$ at the RHS iteratively in the `self-energy' $\Sigma$.
Then, by deriving Eq.~\re{FORMALGEQ} with respect to $\eta_{a}$ we get
\beqra
&& \partial_{\eta_{a}}\,G_{ab}(k;\,\eta_a,\eta_b)= - \Omega_{ac} G_{cb}( k;\,\eta_a,\eta_b)\nonumber\\
&&\qquad\qquad\qquad+\int_{\eta_b}^{\eta_a} d s\, \Sigma_{ac}( k;\,\eta_a\,,s)\,G_{cb}(k;\,s,\eta_b)\, ,
\label{FORMALTRGGEQ}
\eeqra
which gives the exact ({\it i.e.} non-perturbative) evolution equation for the full propagator, and  will be the starting point for our evaluation of $G_{ab}$.

\section{Factorization and the Crocce-Scoccimarro propagator}
\label{FACTORIZATIONSECTION}

In order to get insight on the content of the exact evolution equation, Eq.~(\ref{FORMALTRGGEQ}), we will consider a perturbative expansion for the propagator and the `self-energy',
\beqra
G_{ab}(k;\,\eta_a,\eta_b) &=& \sum_{n=0}^\infty G_{ab}^{(n)}(k;\,\eta_a,\eta_b)\,,\nonumber\\
\Sigma_{ab}(k;\,\eta_a,\eta_b) &=& \sum_{n=1}^\infty\Sigma_{ab}^{(n)}(k;\,\eta_a,\eta_b)\,,
\label{pertexp}
\eeqra
where, as usual, the index $n$ counts the number of power spectra contained in the $n$-th order contributions to $G_{ab}$ and $\Sigma_{ab}$. Notice that at zeroth order $\Sigma_{ab}$ receives no contribution, while
\beq
G_{ab}^{(0)}(k;\,\eta_a,\eta_b) = g_{ab}(\eta_a-\eta_b)\,.
\eeq
Inserting (\ref{pertexp}) in (\ref{FORMALTRGGEQ}), and equating terms of the same order, we get the evolution equation for the $n$-th order contribution to the full propagator,
\beqra
 \partial_{\eta_{a}}\,G_{ab}^{(n)}(k;\,\eta_a,\eta_b)= &-& \Omega_{ac}\, G_{cb}^{(n)}( k;\,\eta_a,\eta_b)\nonumber\\
&+&\Theta_{n,0} \sum_{j=0}^{n-1} \int_{\eta_b}^{\eta_a} d s\, \Sigma_{ac}^{(n-j)}( k;\,\eta_a\,,s)\,G_{cb}^{(j)}(k;\,s,\eta_b)\,,
\label{TRGPERT}
\eeqra
where $\Theta_{n,0}$ is zero for $n=0$ and one otherwise.

\begin{figure}
\centerline{\includegraphics[width = 8cm,keepaspectratio=true]{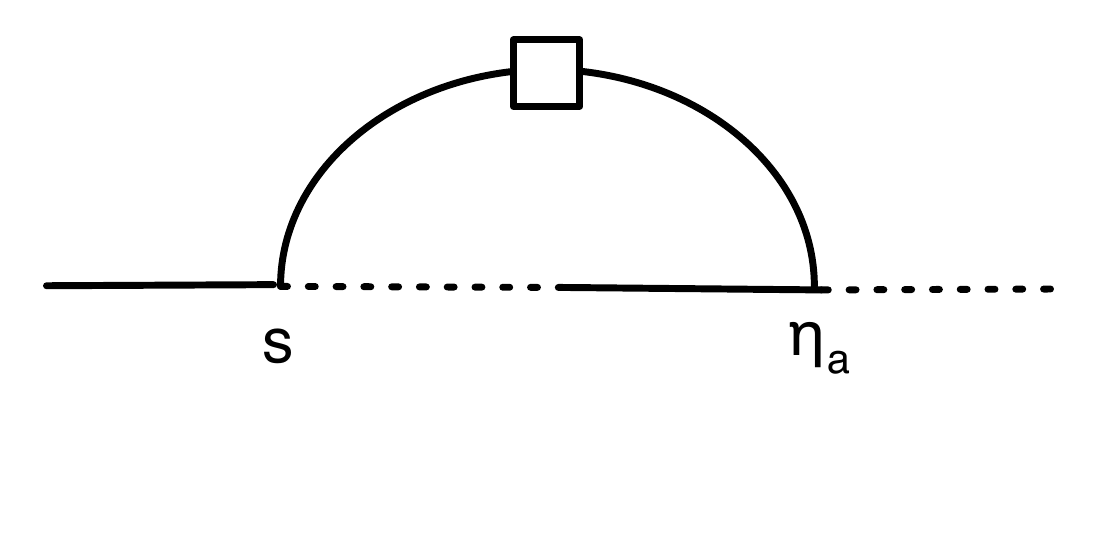}}
\caption{$\Sigma_{ac}^{(1)}$, the 1-loop contribution to the self-energy.}
\label{sig1loop}
\end{figure}

\begin{figure}
\centerline{\includegraphics[width = 15cm,keepaspectratio=true]{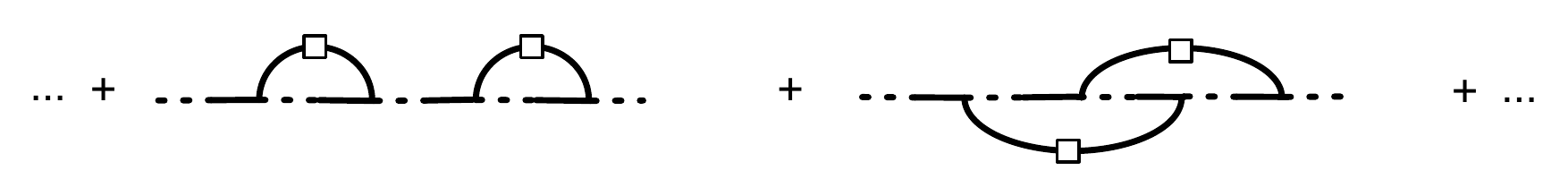}}
\caption{Chain-diagrams.}
\label{FIGCHAIN}
\end{figure}

In the large external momentum limit the leading diagrams contributing to the last line of Eq.~(\ref{TRGPERT}) are the so-called chain-diagrams (see Fig. \ref{FIGCHAIN}) already discussed by CS in \cite{Crocce:2005xz}, all the other contributions being suppressed by inverse powers of $k$. As discussed in detail in \ref{fattorizzazione}, in this limit the sum can be computed analytically, giving the remarkable factorized result
\beqra
 && \sum_{j=0}^{n-1} \int_{\eta_b}^{\eta_a} d s\, \Sigma_{ac}^{(n-j)}( k;\,\eta_a\,,s)\,G_{cb}^{(j)}(k;\,s,\eta_b) \nonumber\\
 && \qquad \stackrel{\mathrm{large}\;k}{\longrightarrow} \;G_{{\bf a}b}^{(n-1)}(k;\,\eta_a,\eta_b) 
\int_{\eta_b}^{\eta_a} d s\, \Sigma_{{\bf a}c}^{(1)}( k;\,\eta_a\,,s)\,u_c\,,
\label{facto}
\eeqra
 where we used the boldface for the index $a$ in the last line to indicate that it is {\em not} summed over. $\Sigma_{ a c}^{(1)}$ represents the 1-loop contribution to the self-energy corresponding to the diagram Fig.~\ref{sig1loop}. In the large momentum limit one gets
\beq
  \int_{\eta_b}^{\eta_a} ds \, \Sigma_{ a c}^{(1)}( k;\,\eta_a \,,s)\,u_c \stackrel{\mathrm{large \; k}}{\longrightarrow} - k^2 \sigma^2 \,e^{\eta_a}(e^{\eta_a}-e^{\eta_b})\,, \;\;(\mathrm{for}\;a=1,2)\,.
\eeq
with
\beq
\sigma^2 \equiv \frac{1}{3} \int d^3 q \frac{P^0(q)}{q^2}\,.
\label{sig2}
\eeq

Summing Eq.~(\ref{TRGPERT}) over $n$, we get the evolution equation for the full propagator in the large momentum limit
\beqra
 \partial_{\eta_{a}}\,G_{ab}(k;\,\eta_a,\eta_b)= &-& \Omega_{ac}\, G_{cb}( k;\,\eta_a,\eta_b)\nonumber\\
&+& G_{{\bf a}b}(k;\,\eta_a,\eta_b) 
\int_{\eta_b}^{\eta_a} d s\, \Sigma_{{\bf a}c}^{(1)}( k;\,\eta_a\,,s)\,u_c\,.
\label{TRGL}
\eeqra
At large $k$ the differential Eq.~\re{TRGL} can be easily integrate in $\eta_a$ and it yields
\beq
G_{ab}(k;\,\eta_a,\eta_b) u_b =  \exp\left(-k^2\sigma^2 \frac{(e^{\eta_a}-e^{\eta_b})^2}{2}\right)\, \;\;(\mathrm{for}\;a=1,2)\,,
\label{resuL}
\eeq
this reproduces the Gaussian decay of the large momentum propagator found by CS in Ref.~\cite{Crocce:2005xz}

In the opposite limit, $k\to 0$, higher order contributions to the propagator are suppressed, and linear perturbation theory is recovered. In order to take into account the first non-linear corrections in this limit, one can truncate the series in (\ref{pertexp}) at $n=1$, {\it i.e.} at 1-loop order. Moreover, we will consider the evolution of the two combinations $G_{a}$, introduced in Eq.~(\ref{Gcont}). Therefore, the relevant term in the sum of Eq.~(\ref{TRGPERT}) in the small $k$ limit is the one for $n=1$, namely,
\beq
\int_{\eta_b}^{\eta_a} d s\, \Sigma_{ac}^{(1)}( k;\,\eta_a\,,s)\,g_{cb}(s -\eta_b) u_b = \int_{\eta_b}^{\eta_a} d s\, \Sigma_{ac}^{(1)}( k;\,\eta_a\,,s)\, u_c\,,
\label{1loop}
\eeq
where we have used the property of the linear propagator,
\beq 
g_{ab}(\eta) u_b=u_a\,.
\label{gu}
\eeq
Modulo terms at least of 2-loop order, the above expression can be rewritten as
\beq
G_{{\bf a}b}(k;\,\eta_a,\eta_b)\, u_b
\int_{\eta_b}^{\eta_a} d s\, \Sigma_{{\bf a}c}^{(1)}( k;\,\eta_a\,,s)\, u_c\,,
\label{smallk}
\eeq
which gives the same equation as the one for large $k$, Eq.~(\ref{TRGL}), contracted by $u_b$ \footnote{If we do not contract by $u_b$, the factorization still holds exactly for the individual components of the propagator at small $k$ if one takes the $\eta_b\to -\infty$ limit. For finite $\eta_b$, the factorization is not exact anymore, but it is anyway a very good approximation.}. It is remarkable that the same factorization holds in the two limits of large and small $k$.

For any $k$ value the 1-loop contribution to the self-energy reads 
\beqra
&&\Sigma_{ac}^{(1)}( k;\,\eta_a\,,s)=\nonumber\\
&& 4 e^{\eta_a+s} \int d^3 q\, \gamma_{ade}(\bk,-\bq,\bq-\bk)P^0(q)u_du_f g_{eh}(\eta_a-s) \gamma_{hfc}(\bk-\bq,\bq,-\bk)\,,\nonumber\\
&&
\label{s1loop}
\eeqra
and allows us to compute the propagator both in the small and large momentum limits with the evolution Eq.~(\ref{TRGL}). Notice that the PS appearing in Eq.~(\ref{s1loop}) is the linear one.

The above discussion clarifies the comparison between the present approach and the one in \cite{Crocce:2005xz}. Indeed, both give the same result, Eq.~(\ref{resuL}) for $k\to\infty$, and  both reproduce the 1-loop propagator for $k\to0$. For intermediate $k's$ the two approaches give different prescriptions to interpolate between their common limits. In the present section, the interpolation is based on two approximations.
 First, we factorize the integral of Eq.~(\ref{FORMALTRGGEQ}) as
\beqra
&&\int_{\eta_b}^{\eta_a} d s\, \Sigma_{ac}( k;\,\eta_a\,,s)\,G_{cb}(k;\,s,\eta_b) \simeq G_{{\bf a}b}(k;\,\eta_a,\eta_b) 
\int_{\eta_b}^{\eta_a} d s\, \tilde{\Sigma}_{{\bf a}c}( k;\,\eta_a\,,s)\,u_c\,,\nonumber\\
&&
\label{factoApproxim}
\eeqra
where $\tilde{\Sigma}_{{\bf a}c}$ refers to the approximation chosen for the two-point 1PI function. Second, we consider the purely 1-loop `self-energy' in the factorized expression at the RHS above 
\beq
\tilde{\Sigma}_{{\bf a}c}( k;\,\eta_a\,,s) \simeq \Sigma_{{\bf a}c}^{(1)}( k;\,\eta_a\,,s)\,.
\label{factoApproxim1loop}
\eeq
These are the two approximations needed to pass from the exact equation (\ref{FORMALTRGGEQ}) to the approximated one, Eq.~(\ref{TRGL}), which we will solve for any $k$. On the other hand, in \cite{Crocce:2005xz} the interpolation is obtained by advocating an exponentiation procedure. We can directly check the differences between the two procedures by taking the $\eta$ derivative of the propagator given in Eq.~(41) of  \cite{Crocce:2005xz}. In the limit $\eta_i\to - \infty$, the comparison simplifies considerably, and one finds an evolution equation given by Eq.~(\ref{TRGL}) plus an extra term at the RHS, given by
\beq
 \Omega_{{\bf a} b}  \left[\exp \left( {G^{(1)}_{bc}u_c} \right)-{G^{(1)}_{bc}u_c} \exp \left( {G^{(1)}_{{\bf a}d}u_d} \right)  \right]\,,
\eeq
where 
\beq G^{(1)}_{ab}(k; \eta,\eta_i) = \int_{\eta_i}^\eta ds\int_{\eta_{i}}^s ds^\prime g_{ac}(\eta-s) \Sigma^{(1)}_{cd}(k;s,s^\prime) g_{db}(s^\prime-\eta_i)\,,
\eeq
 is the 1-loop contribution to the propagator, and $\Omega_{ab}$ has been defined in Eq.~(\ref{bigomega}).
The extra term is at most of two-loop order and it vanishes both in the large and small momentum limits, as it should. Therefore, in the rest of this paper, we will refer to the propagator obtained by solving Eq.~(\ref{FORMALTRGGEQ})  in the double approximation (\ref{factoApproxim}) and (\ref{factoApproxim1loop}) as the CS propagator, even though it differs from the one of ref.~\cite{Crocce:2005xz} by subleading terms at intermediate $k$. We stress again that the difference between the propagator computed along the lines described in this section and the one in ~\cite{Crocce:2005xz} is entirely due to the different interpolation procedures, while the classes of diagrams resummed are exactly the same.

In the following we will go beyond the approximation of this section in a consistent way. As mentioned above, we will keep the factorized form of the equation for any $k$, but will consider new contributions to the `self-energy', therefore improving over the approximation of Eq.~ (\ref{factoApproxim1loop}). As we will show, this corresponds to resumming a larger class of diagrams than just the infinite chains considered by CS.

\section{Extended factorization: the renormalized chain-diagrams}
\label{FACTORIZATIONSECTIONPS}
\begin{figure}
\centerline{\includegraphics[width = 8cm,keepaspectratio=true]{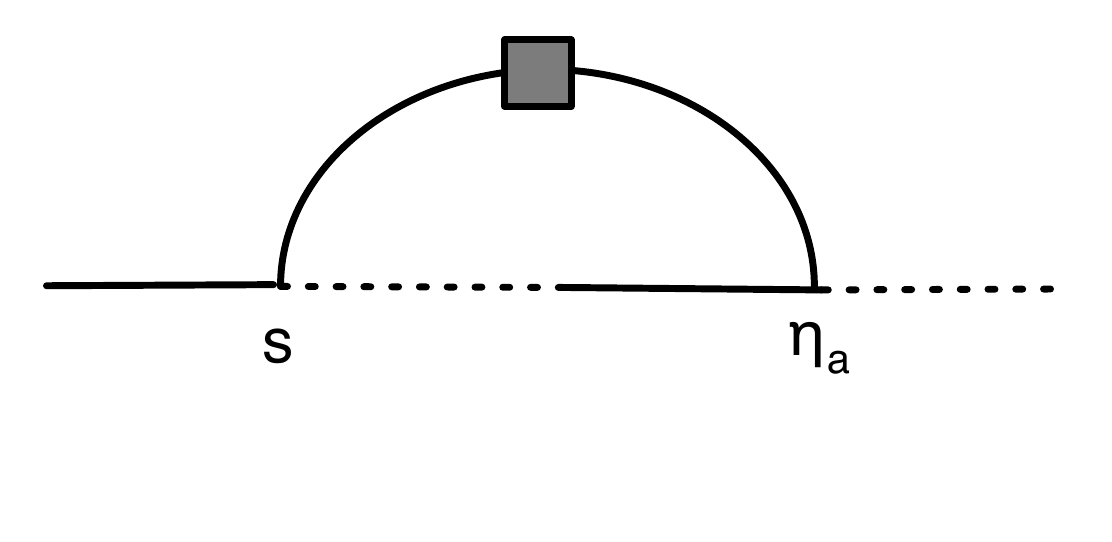}}
\caption{$\Sigma_{ac}^\mathrm{PSnl}( k;\,\eta_a\,,s)$: 1-loop self-energy with the insertion of the non linear PS.}
\label{SIGMA1}
\end{figure}

The large-$k$ factorization property, Eq.~(\ref{facto}), holds for a more general class of diagrams than the `chain'-ones discussed by CS. This is the main result of this paper, and is proved in \ref{fattorizzazione}. There, we show that by replacing all the linear power spectra, $P_{ab}^0(q)=P^0(q) u_a u_b$, appearing in the chain-diagrams by a -- for now --  generic {\it non-linear} PS, of the form $P^{\mathrm{nl}}_{ab}(q; s_a,s_b)$, one still gets Eq.~(\ref{facto}) in the large-$k$ limit, where now, 
\beq
\tilde{\Sigma}_{{\bf a}c}( k;\,\eta_a\,,s) \simeq \Sigma_{{\bf a}c}^{\mathrm{PSnl}}( k;\,\eta_a\,,s)\,.
\label{psnlapp}
\eeq
$ \Sigma_{{\bf a}c}^{\mathrm{PSnl}}$ is obtained from the 1-loop self-energy by replacing  $P_{ab}^0$ by $P^{\mathrm{nl}}_{ab}$, as indicated in the diagram in Fig.~\ref{SIGMA1}, 
and has the large-momentum limit
\beq
\int_{\eta_b}^{\eta_a} ds\,\Sigma_{{\bf a}c}^{\mathrm{PSnl}}( k;\,\eta_a\,,s) u_c \stackrel{\mathrm{large} \; k}{\longrightarrow}   \left(\frac{-k^2}{3}\right) \int_{\eta_b}^{\eta_a}ds \, e^{\eta_a+s} \int d^3 q \frac{P^{\mathrm{nl}}_{22}(q;\eta_a, s)}{q^2} \,.
\label{largenl}
\eeq
The effect of the inclusion of these subleading corrections is clear: at large momenta the propagator still decays exponentially, but with the decay law of Eq.~(\ref{resuL}) replaced by 
\beq
G_{ab}(k;\,\eta_a,\eta_b) u_b  =  \exp\left(-k^2\sigma_\mathrm{nl}^2(\eta_a,\eta_b) \frac{(e^{\eta_a}-e^{\eta_b})^2}{2}\right)\, \;\;(\mathrm{for}\;a=1,2)\,,
\label{resutNL}
\eeq
where
\beq
\sigma_\mathrm{nl}^2(\eta_a,\eta_b) \frac{(e^{\eta_a}-e^{\eta_b})^2}{2} \equiv  \frac{1}{3}\int_{\eta_b}^{\eta_a} ds_1\int_{\eta_b}^{s_1} ds_2 \,\e^{s_1+s_2} \int d^3 q\frac{P^{\mathrm{nl}}_{22}(q; s_1,s_2)}{q^2}\,.
\label{sigt}
\eeq
Notice that only the ``22" ({\it i.e.} velocity-velocity) component of the PS appears in the exponential above. Since it is known that this component receives negative corrections at the non-linear level, (see, for instance, \cite{Pietroni:2008jx,Hiramatsu:2009ki}), we expect that the improved propagator will be enhanced w.r.t the CS one at large $k$.

In the opposite limit, $k \rightarrow 0$, we are no longer guaranteed that Eq.~(\ref{psnlapp}) is still a good approximation, and that the contributions to the $\tilde{\Sigma}_{ac}$ obtained by replacing the linear PS with the non-linear one are the only leading ones. In other words we consider again the n-order contribution expressed in Eq.~\re{TRGPERT} where now the upper indices count the number of non-linear PS in a given contribution to $\Sigma$ and $G$. In this context, in the small-$k$ limit, again the relevant term in the sum of Eq.~(\ref{TRGPERT}) is the one for $n=1$. Following the same argument of the previous section the non linear part of Eq.~\re{TRGPERT} becomes
\beq
G_{{\bf a}b}(k;\,\eta_a,\eta_b)\, u_b
\int_{\eta_b}^{\eta_a} d s\, \Sigma_{{\bf a}c}^{\mathrm{PSnl}}( k;\,\eta_a\,,s)\, u_c\,,
\label{smallkPSnl}
\eeq
getting the factorization property in the small-$k$ limit. At this level we avoid any double counting problem, however the application of this perturbative criterion must be analyzed carefully by comparing the results with the standard perturbative computation. In this respect, a powerful guiding criterium is the requirement that linear theory is recovered for small momenta. Indeed, from the exact equations (\ref{GEXPANSION}) and (\ref{FORMALGEQ}), one concludes that, in order to have $G_{ab}(k; \eta_a,\eta_b) \to g_{ab}(\eta_a-\eta_b)$ as $k\to0$ the `self-energy' has to vanish in this limit
\beq
\Sigma_{ab}(k;\,\eta_a,\,s) \stackrel{k\to 0 }{\longrightarrow} 0\,,
\label{limk0}
\eeq
therefore we should also have $\Sigma_{a c}^{\mathrm{PSnl}}( k;\,\eta_a\,,s)$ vanishing for $k\to 0$.
In the approximation (\ref{smallkPSnl}) this is not automatically realized. Indeed, one finds
\beqra
&&\int_{\eta_b}^{\eta_a} d s\,  \Sigma_{{ a}c}^{\mathrm{PSnl}}( k;\,\eta_a\,,s)\, u_c\,\stackrel{k \to 0}{\longrightarrow} \nonumber\\
&& \frac{1}{3} \delta_{a1} \int_{\eta_b}^{\eta_a} d s\, e^{\eta_{a}+s}\,
\int d^3 q \left[ g_{2d}(\eta_{a} -s)P^\mathrm{nl}_{1d}(q;\eta_a, s)-g_{1d}(\eta_{a} -s)P^\mathrm{nl}_{2d}(q;\eta_a, s)\right].
 \nonumber
\\
\label{1loopPSzeroK}
\eeqra

Of course, if one puts back the linear PS, $P^0_{ab}=P^0u_a u_b$, in place of $P^\mathrm{nl}_{ab}$ in the expression above, one recovers the 1-loop self-energy, which vanishes in the $k\to 0$ limit as can be directly checked from Eq.~(\ref{1loopPSzeroK}), using Eq.~(\ref{gu}) (the first non-vanishing contribution goes as $k^2$).

\begin{figure}
\centerline{\includegraphics[width = 15cm,keepaspectratio=true]{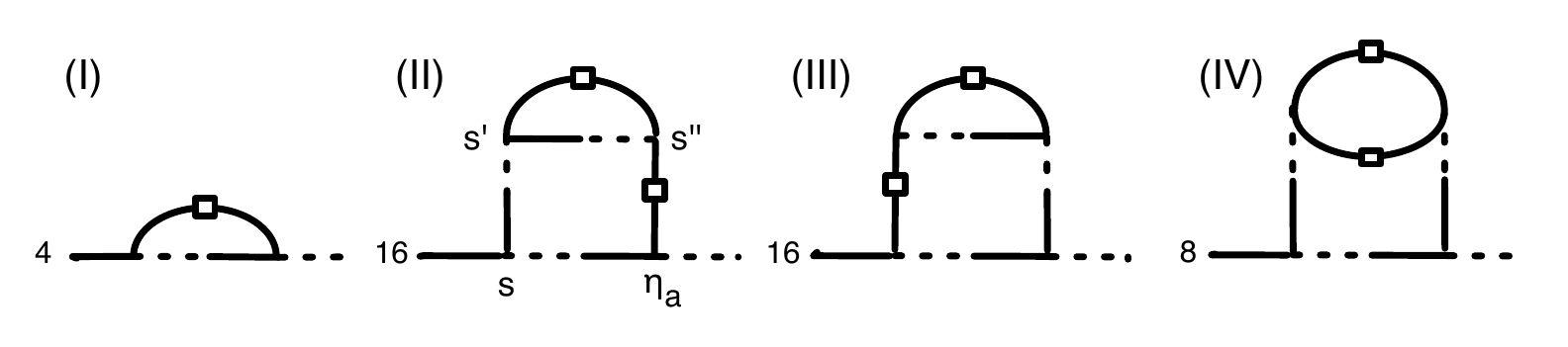}}
\caption{$\Sigma_{ac}^\mathrm{PS1l}( k;\,\eta_a\,,s)$: the self-energy with the insertion of the PS up to 1-loop order.}
\label{FIGSIGMAPS1LOOP}
\end{figure}

\begin{figure}
\centerline{\includegraphics[width = 15cm,keepaspectratio=true]{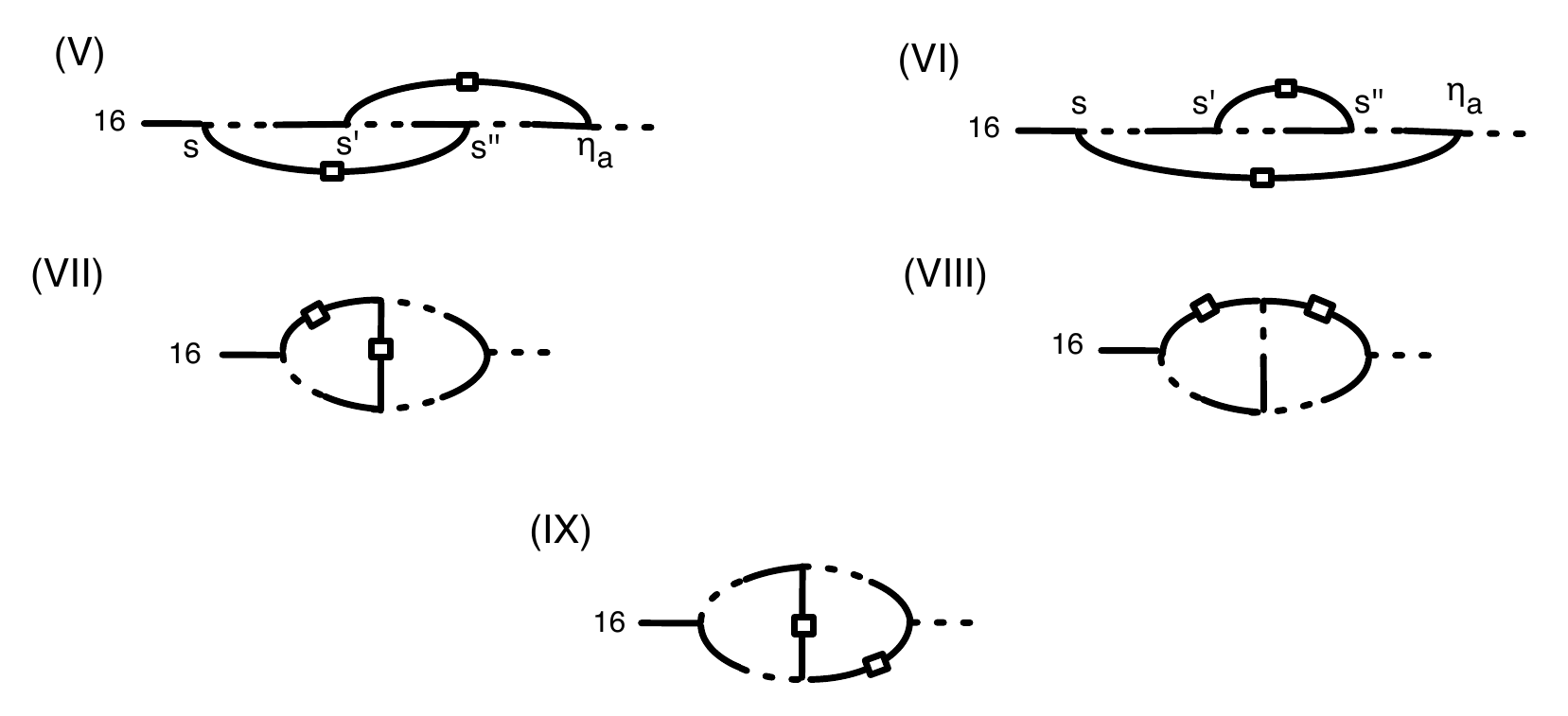}}
\caption{$\Sigma_{ ac}^{\mathrm{2l_{rest}}}$: the 2-loop contributions to the self-energy not included in $\Sigma_{ac}^\mathrm{PS1l}$.}
\label{FIGSIGMA2LOOP}
\end{figure}

On the other hand, moving a step further and using the 1-loop result for $P^\mathrm{nl}_{ab}$, namely, including the diagrams of Fig.~\ref{FIGSIGMAPS1LOOP} in the computation of the self-energy, one finds a non-vanishing limit for $k\to0$. Indeed, one can check that the contributions from the remaining 2-loop diagrams listed in Fig.~\ref{FIGSIGMA2LOOP} exactly cancel those of Fig.~\ref{FIGSIGMAPS1LOOP}, recovering in this way the physical requirement of Eq.~(\ref{limk0}). 

At large $k$ the contributions V and VI in Fig.~\ref{FIGSIGMA2LOOP} would give rise to  chain-diagrams for the propagator with the insertion of a linear PS. These contributions are already taken into account by diagram I in Fig.~\ref{FIGSIGMAPS1LOOP}, so,  in order to avoid double counting, we do not have to include them. The remaining diagrams, VII-IX, are subdominant at large $k$ w.r.t. the chain-diagrams. On the other hand, at small $k$, all the diagrams in Fig.~\ref{FIGSIGMA2LOOP} are essential in order to recover linear theory. Therefore, an approximation to $\tilde{\Sigma}_{ ac}$ giving the `1-loop renormalized' chain-diagrams ({\it i.e.} the chain-diagrams with the 1-loop PS replacing the linear one) in the large $k$ limit, and recovering linear theory for $k\to 0$, is given by
\beqra 
\tilde{\Sigma}_{ ac}( k;\,\eta_a\,,s) &\simeq& \Sigma_{ ac}^{\mathrm{PS1l}}( k;\,\eta_a\,,s) + \lim_{k\to 0} 
\Sigma_{ ac}^{\mathrm{2l_{rest}}}( k;\,\eta_a\,,s) \nonumber\\
&=& \Sigma_{ ac}^{\mathrm{PS1l}}( k;\,\eta_a\,,s) - \lim_{k\to 0} 
\Sigma_{ ac}^{\mathrm{PS1l}}( k;\,\eta_a\,,s)\,,
\label{SigmR1l}
\eeqra
where $\Sigma_{ ac}^{\mathrm{PS1l}}$ and $\Sigma_{ ac}^{\mathrm{2l_{rest}}}$ are the contributions to the self-energy computed from the diagrams of figs.~\ref{FIGSIGMAPS1LOOP} and \ref{FIGSIGMA2LOOP}, respectively. The $\Sigma_{ ac}^{\mathrm{2l_{rest}}}( k;\,\eta_a\,,s)$ term gives also contributions of order $k^2$ when $k$ approaches to zero, that we do not include with the prescription in \re{SigmR1l}. Therefore, our resummed propagator reproduces the linear one in the  $k\to 0$ limit, with the $O(k^{2})$ terms given by the 1-loop one plus some, but not all, of the 2-loop and higher contributions.

\begin{figure}
\centerline{\includegraphics[width = 15cm,keepaspectratio=true]{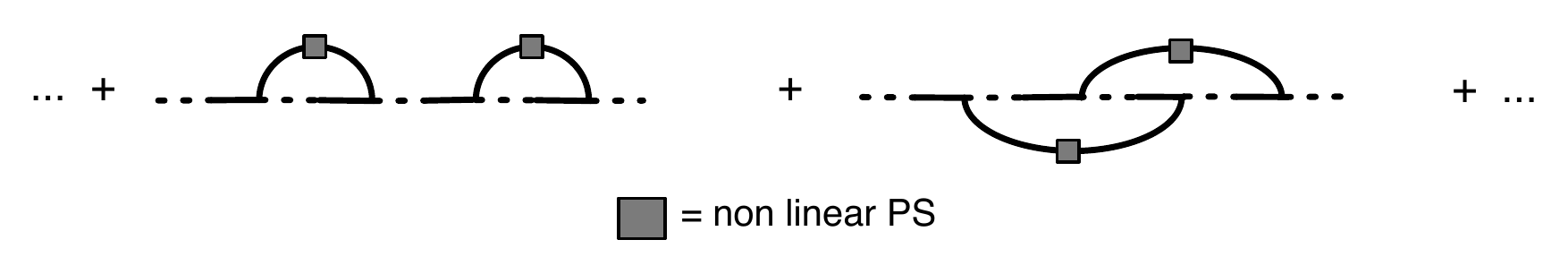}}
\caption{Renormalized chain-diagrams.}
\label{FIGCHAINPS}
\end{figure}

Finally we point out that using the 1-loop PS corresponds to compute Eq.~\re{TRGPERT} for $n=1$ and for $n=2$ with the usual non-renormalized Feynman diagrams. Therefore one can check that the approximation \re{smallkPSnl} holds in this case. Indeed computing the non linear term of Eq.~\re{TRGPERT} at 2-loop level one gets
\beq
\int_{\eta_b}^{\eta_a} d s\, (\Sigma_{ac}^{(1)}( k;\,\eta_a\,,s)\,G^{(1)}_{cb}(k;\, s\,,\eta_b)+\Sigma_{ac}^{(2)}( k;\,\eta_a\,,s)\,g_{cb}(s -\eta_b)) u_b\, .
\label{facto2loop}
\eeq
In Eq.~\re{facto2loop} the first term goes as $k^4$ when $k$ approaches to zero and results subdominant w.r.t. the second one (that goes as $k^2$ ). This allows to consider just the second term and to advocate the procedure involved in the previous section to achieve the factorization property also at 2-loop order. This argument justifies our factorization procedure given by expression \re{smallkPSnl} also for small values of $k$.


\begin{figure}
\centerline{\includegraphics[width = 15cm,keepaspectratio=true]{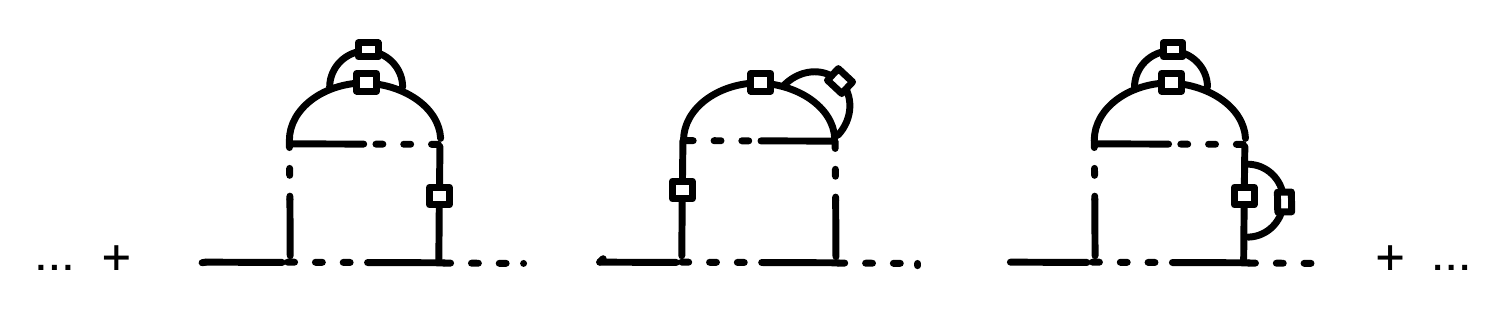}}
\caption{Some of the contributions included in $\tilde{\Sigma}_{ab}$ when the non-linear PS is given by the one computed via the TRG.}
\label{FIGPSTRG}
\end{figure}

A further extension of the resummation program is to use as non-linear PS in the computation of $\Sigma_{ac}^\mathrm{PSnl}$ the one computed with the  Time Renormalization Group (TRG) approach introduced in \cite{Pietroni:2008jx} and briefly reviewed in \ref{PHIPHIAPP}. As discussed in \cite{Pietroni:2008jx}, the TRG equations, truncated at the bispectrum level, incorporate perturbative corrections in which the PS lines are iteratively replaced by their 1-loop corrections. This procedure resums perturbative contributions at all orders, of which some 3 and 4-loop examples are listed in Fig.~\ref{FIGPSTRG}. The TRG approach is able to reproduce the non-linear PS at $z=0$ up to $k \alt 0.2\; \mathrm{h/Mpc}$ (that is, in the BAO region) at the few percent level \cite{Pietroni:2008jx, Carlson:2009it}. 

The TRG, as it is, gives the non-linear PS computed at equal times, that is $P^\mathrm{TRG}_{ab}(q; \eta, \eta)$. On the other hand, in the computation of the self-energy, we need the PS computed at different times, see Eqs.~(\ref{largenl}, \ref{1loopPSzeroK}). The relation between the equal-times PS and the one computed at different times can be read from Eq.~(\ref{PSGNG}) where we now take the initial time $\eta=0$ to be any generic time, $\eta=s$. If $s$ corresponds to a low redshift, non-linearities and non-gaussianities cannot be neglected. They are encoded in the non-linear initial PS, that we will compute with the TRG, and in the irreducible function $\Phi_{cd}(k;\eta_a,s)$, which gives a subdominant contribution that we will compute in PT.
When dealing with $P^\mathrm{TRG}_{ab}(q; \eta_a, s)$ we will therefore replace it by
\beqra
P^\mathrm{TRG}_{ab}(q; \eta_a, s) &\simeq& G^{CS}_{ac}(q; \eta_a,s)P^\mathrm{TRG}_{cb}(q; s, s)\nonumber\\
&&+\int_{s}^{\eta_a} e^{s_{1}}\,\int d^{3}q\,g_{ac}(\eta_{a}-s_{1})\gamma_{cde}(\bk,\,-\bq,\,\bq-\bk)\,\nonumber\\
&&\times\,g_{df}(s_{1}-s)\,g_{eg}(s_{1}-s)\,B_{bfg}(\bk,\,-\bq,\,\bq-\bk; s) \, ,
\label{presc}
\eeqra
where, for practical reasons, we have replaced the full propagator with, $G^{CS}_{ac}$, the propagator computed \`a la Crocce-Scoccimarro according to the approximation discussed in Section \ref{FACTORIZATIONSECTION}, and $B$ is the tree-level Bispectrum computed via perturbation theory.

As in the computation of the $\Sigma_{ ac}^{\mathrm{PS1l}}$, we find a non-vanishing  $k\to 0$ limit for the self-energy computed with  $P^\mathrm{TRG}$. We follow the same arguments discussed above and, also in this case, we incorporate the relevant corrections in this limit by using the prescription 
\beqra 
\tilde{\Sigma}_{ ac}( k;\,\eta_a\,,s) \simeq \Sigma_{ ac}^{P^\mathrm{TRG}}( k;\,\eta_a\,,s) - \lim_{k\to 0} \Sigma_{ ac}^{P^\mathrm{TRG}}( k;\,\eta_a\,,s)\,.
\label{SigmRTRG}
\eeqra

To summarize the results of this section, the evolution equation for the propagator including next-to-leading corrections (the renormalized chain-diagrams plus the contributions needed to recover the proper $k \to 0$ limit) is given by
\beqra
 \partial_{\eta_{a}}\,G_{ab}(k;\,\eta_a,\eta_b)= &-& \Omega_{ac}\, G_{cb}( k;\,\eta_a,\eta_b)\nonumber\\
&+& G_{{\bf a}b}(k;\,\eta_a,\eta_b) 
\int_{\eta_b}^{\eta_a} d s\, \tilde{\Sigma}_{{\bf a}c}( k;\,\eta_a\,,s)\,u_c\,,
\label{TRGREN}
\eeqra
which we will solve for any scale $k$ using the two approximations, Eqs.~(\ref{SigmR1l}) and (\ref{SigmRTRG}), for the self-energy $\tilde{\Sigma}_{ac}$.

\section{Results}
\label{results}

\begin{figure}
\centerline{\includegraphics[width = 15cm,keepaspectratio=true]{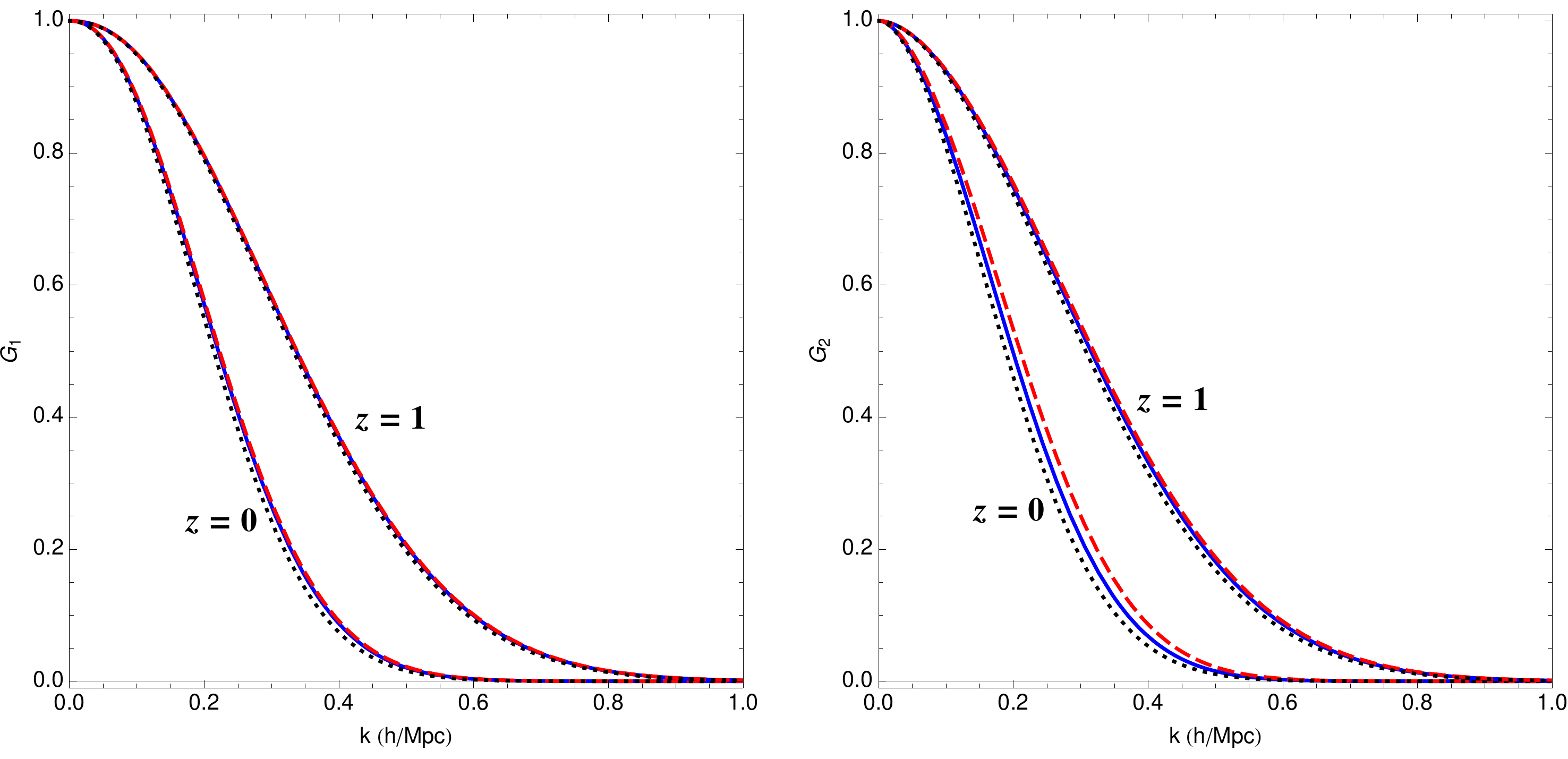}}
\caption{The density (left panel) and velocity (right panel) propagators at redshifts $z=0$ and $z=1$. The black-dotted lines are the propagator computed in the CS approximation. The purple-dashed lines are obtained using the 1-loop PS in $\tilde{\Sigma}$, while the blue-continuous lines are obtained by using the PS from the TRG.}
\label{G1G2}
\end{figure}

We investigate a $\Lambda$CDM cosmology close to the best-fit model ($\Omega_{m}=0.25$, $\Omega_{b}h^2=0.0224$, $h=0.72$, $n=0.97$ and $\sigma_{8}=0.8$). The initial time,  $\eta=0$, is taken to correspond to the physical redshift  $z_{in}=100$. At $\eta=0$ we set the initial conditions for the evolution equation (\ref{TRGREN}) and for the TRG equations needed to compute the PS (see \ref{PHIPHIAPP}). We set the initial conditions for the PS by matching it with the linear PS obtained by the CAMB code \cite{Lewis:1999bs}. For the propagator, we use the explicit expression given in Eq.~(\ref{proplin}). The integration of the TRG equations requires also initial values for the bispectra. We set them to zero,
 {\it i.e.}, we neglect all non-Gaussianities generated at redshifts higher than $z=100$. 
 
In Fig.~\ref{G1G2} we plot our results for the propagators $G_a(k;\ \eta_a,0)$ defined in Eq.~(\ref{Gcont})
($G_{1}$ in the left panel, $G_{2}$ in the right panel), computed at final times $\eta(z)$ corresponding to redshift $z=0$ and $z=1$. 
The dotted black lines correspond to the Crocce-Scoccimarro result, {\it i.e.} to the integration of Eq.~(\ref{TRGL}), where the linear PS has been used to compute $\Sigma^{(1)}_{ac}$. The dashed red lines are obtained by using the 1-loop approximation for the non-linear PS in $\tilde{\Sigma}_{ac}$, while the continuous blue lines are obtained by using the TRG PS.

\begin{figure}
\centerline{\includegraphics[width = 15cm,keepaspectratio=true]{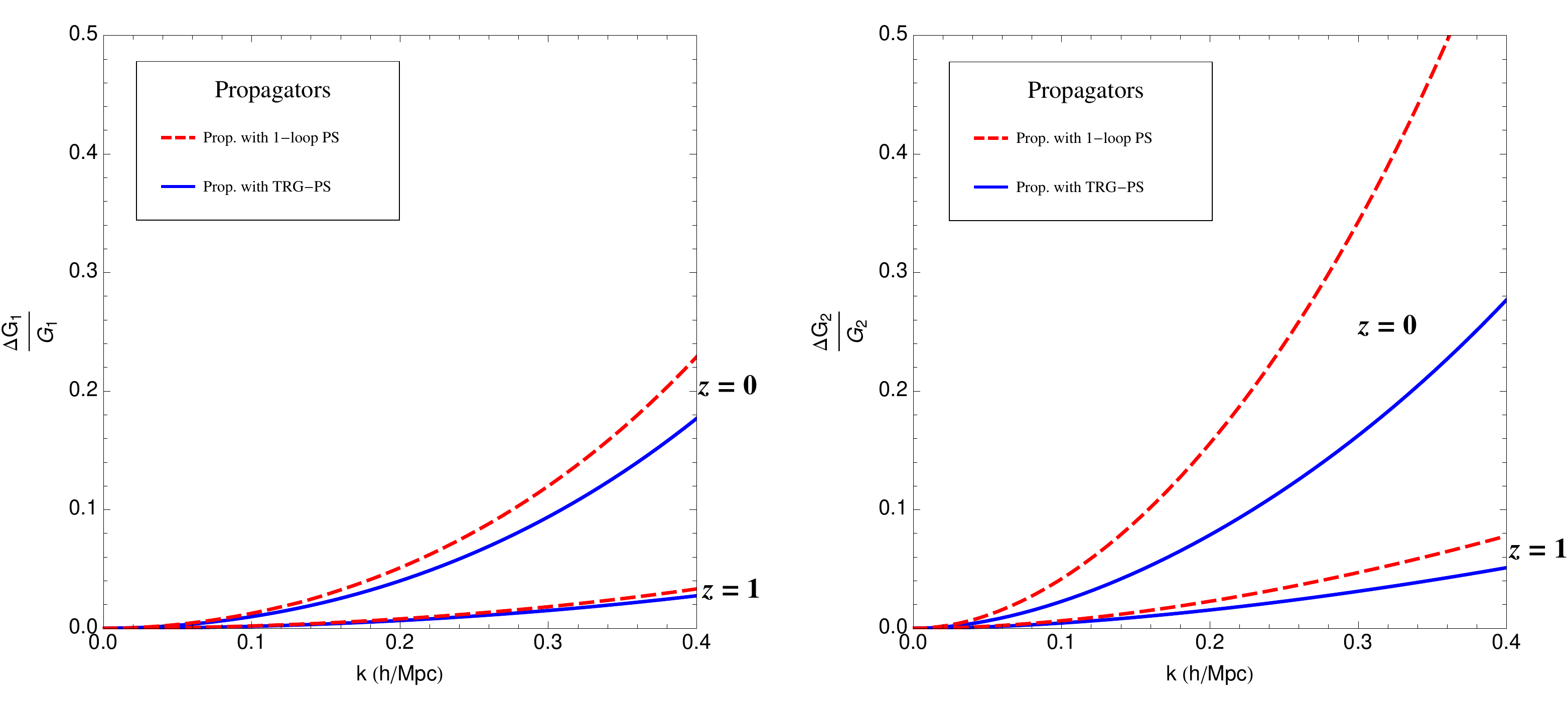}}
\caption{Relative differences between the improved propagators and the one obtained in the CS approximation. Line-codes as in Fig.~\ref{G1G2} }
\label{DiffPercG1G2}
\end{figure}

In Fig.~\ref{DiffPercG1G2} we plot the relative difference between the propagators computed with the two different approximations for the non-linear PS and the one computed in the Crocce-Scoccimarro approximation.  

\begin{figure}
\centerline{\includegraphics[width = 15cm,keepaspectratio=true]{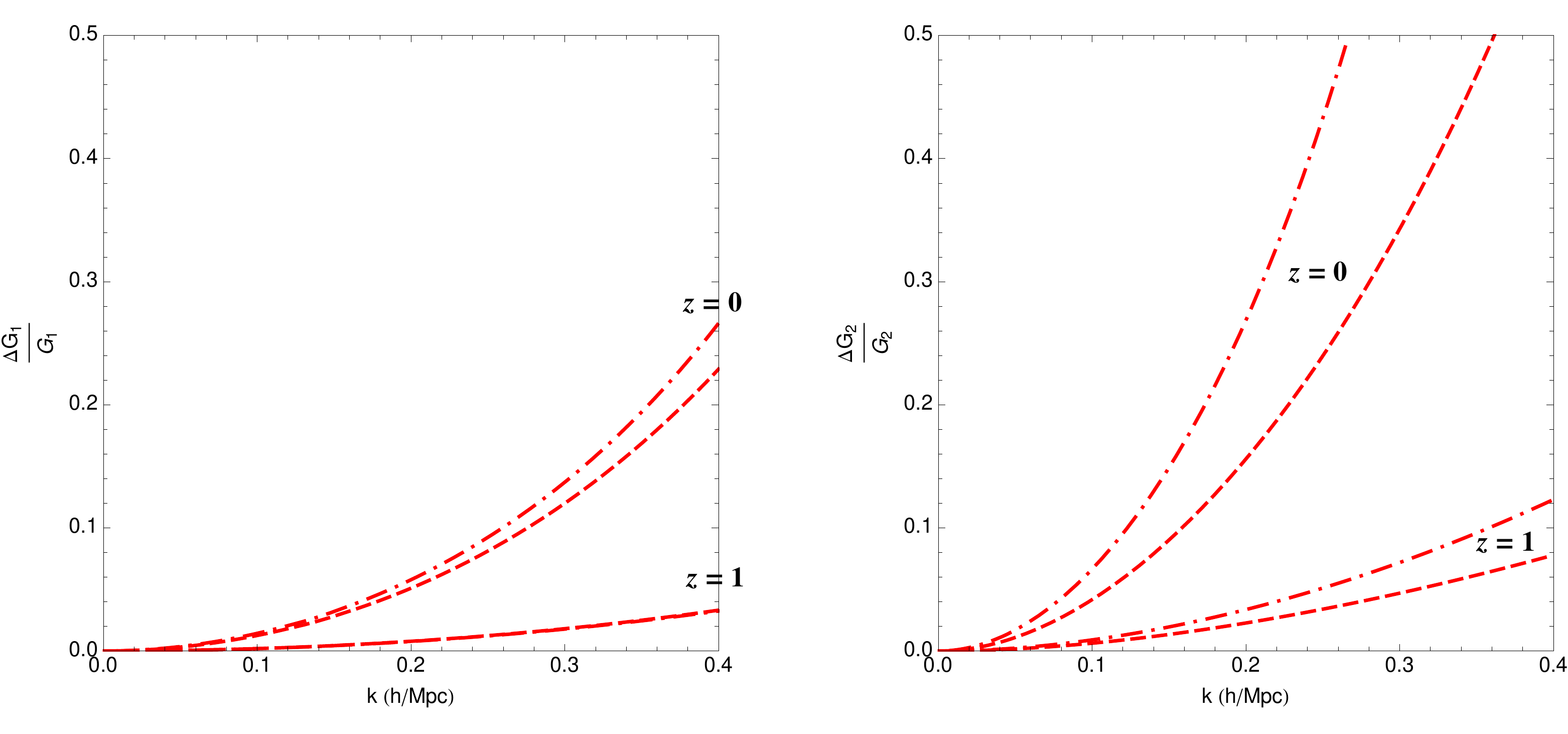}}
\caption{The dependence on the UV-cutoff of the improved propagator obtained by using the 1-loop approximation for the non-linear PS. Dashed lines are obtained using $q_{uv}=1$ h/Mpc$^{-1}$, while dash-dotted lines are obtained with $q_{uv}=2$ h/Mpc$^{-1}$.}
\label{1loopCutoff}
\end{figure}

As a general trend, the effect of the inclusion of the new class of diagrams considered in this paper leads to a weaker damping of the propagators at intermediate and large $k$'s, compared to the one obtained considering only the chain-diagrams of CS. The effect is stronger for the velocity propagator than for the density one.

The computation with the 1-loop PS suffers from an intrinsic uncertainty, due to the dependence on the (UV) momentum cutoff employed in the loop integrals. This is an unavoidabe limitation of the 1-loop approximation, due to the fact that the 1-loop PS takes unphysical negative values at large $q$'s, especially when the two time arguments are very different. In 
Fig.~\ref{1loopCutoff}, we show the cutoff dependence by plotting the same quantities as in Fig.~\ref{DiffPercG1G2}, computed using the 1-loop approximation for the PS and UV cutoffs $q_{\mathrm{max}}=1\; \mathrm{and}\;\,2 \,{\mathrm {h/Mpc}}$. The cutoff dependence is quite strong for the $G_2$ propagator, showing that its computation using the 1-loop PS is clearly unreliable at low redshift.
On the other hand, the results for $G_{1,2}$ obtained using the PS computed with the TRG, which is always positive, do not exhibit UV problems. In this paper we use always the $q_{\mathrm{max}}=1\,{\mathrm {h/Mpc}}$ cutoff limiting in this way the effects due to the unphysical behavior of the 1-loop power spectrum approximation.

\begin{figure}
\centerline{\includegraphics[width = 15cm,keepaspectratio=true]{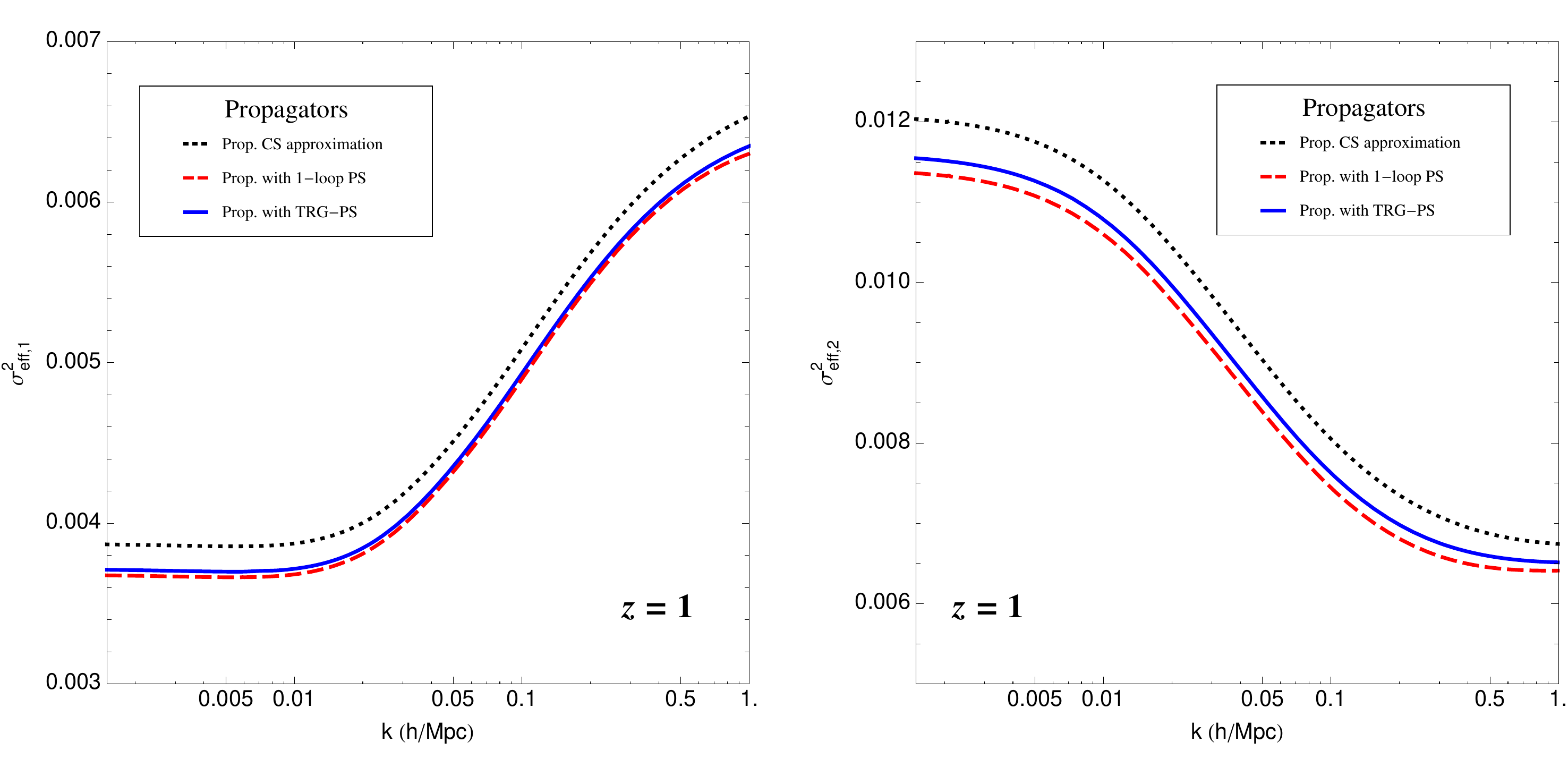}}
\caption{$\sigma_{eff,a}^2$ at $z=1$ computed from $G_{a}$ (see Eq.~\ref{SIGMA2EFF}).  Line-codes as in Fig.~(\ref{G1G2}).}
\label{SIGMAG1G2z1}
\end{figure}

\begin{figure}
\centerline{\includegraphics[width = 15cm,keepaspectratio=true]{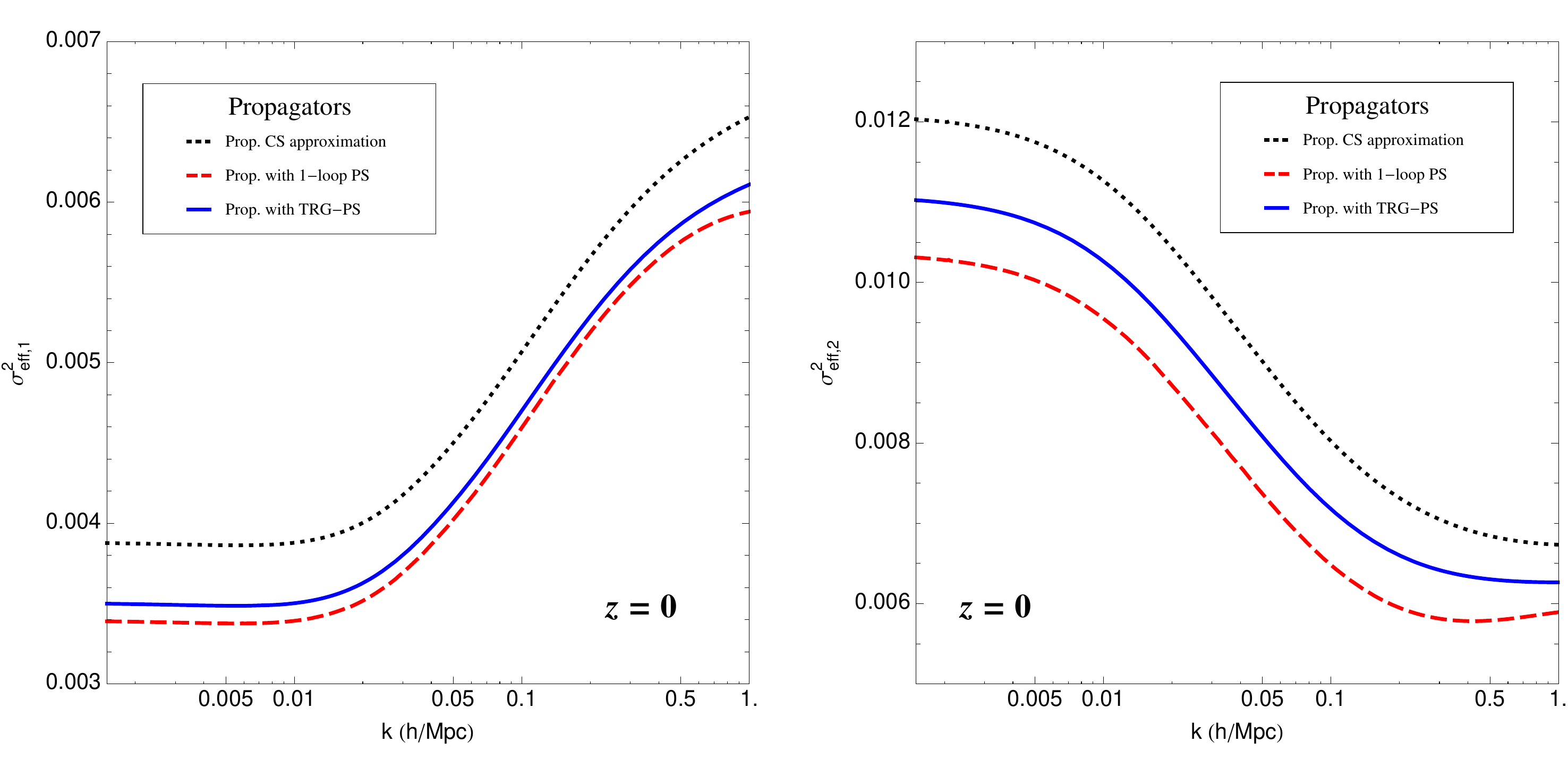}}
\caption{The same as Fig.~\ref{SIGMAG1G2z1} but at $z=0$.}
\label{SIGMAG1G2z0}
\end{figure}

An alternative way to show the effects of our improved approximation is to define an effective velocity dispersion,
\beq
\sigma^2_{eff,a}(k;\eta,0)=\frac{-2}{k^2 (e^{\eta}-1)^2}\ln (G_{a}(k; \eta, 0))\, ,
\label{SIGMA2EFF}
\eeq
which, in the high-$k$ limit, reduces to $\sigma_\mathrm{nl}^2(\eta,0)$ defined in  Eq.~\re{resutNL}. 
In figs. \ref{SIGMAG1G2z1}, \ref{SIGMAG1G2z0}, we plot  $\sigma^{2}_{eff,1}$ (on the left) and $\sigma^{2}_{eff,2}$ (on the right) as a function of  $k$: the line code is the same as in figs.~\ref{G1G2} and \ref{DiffPercG1G2}.

\begin{figure}
\centerline{\includegraphics[width = 15cm,keepaspectratio=true]{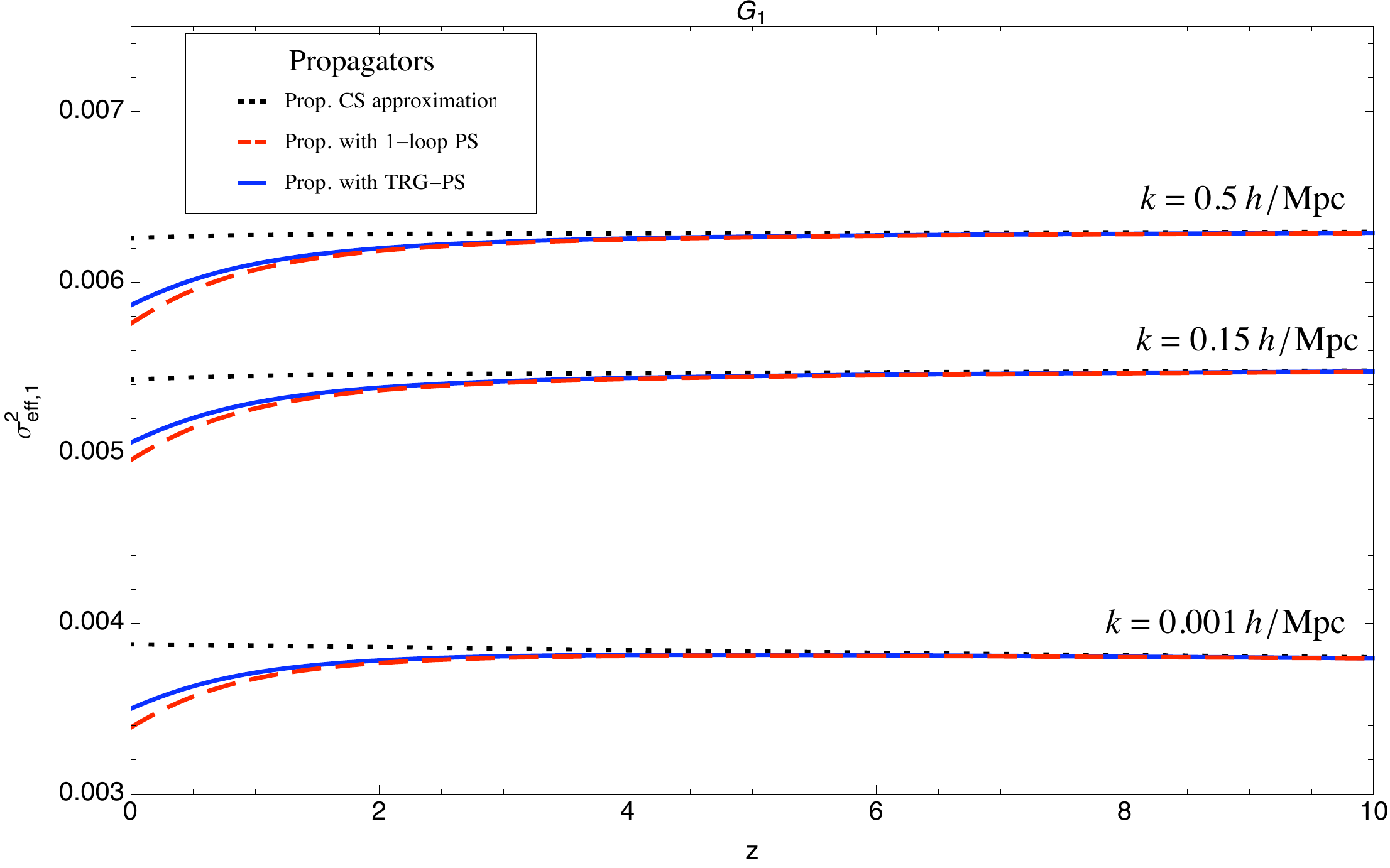}}
\caption{$\sigma_{eff,1}^2(z)$ as a function of redshift for three different momentum scales. Line-codes as in Fig.~\ref{G1G2}. }
\label{SIGMAG1Z}
\end{figure}

\begin{figure}
\centerline{\includegraphics[width = 15cm,keepaspectratio=true]{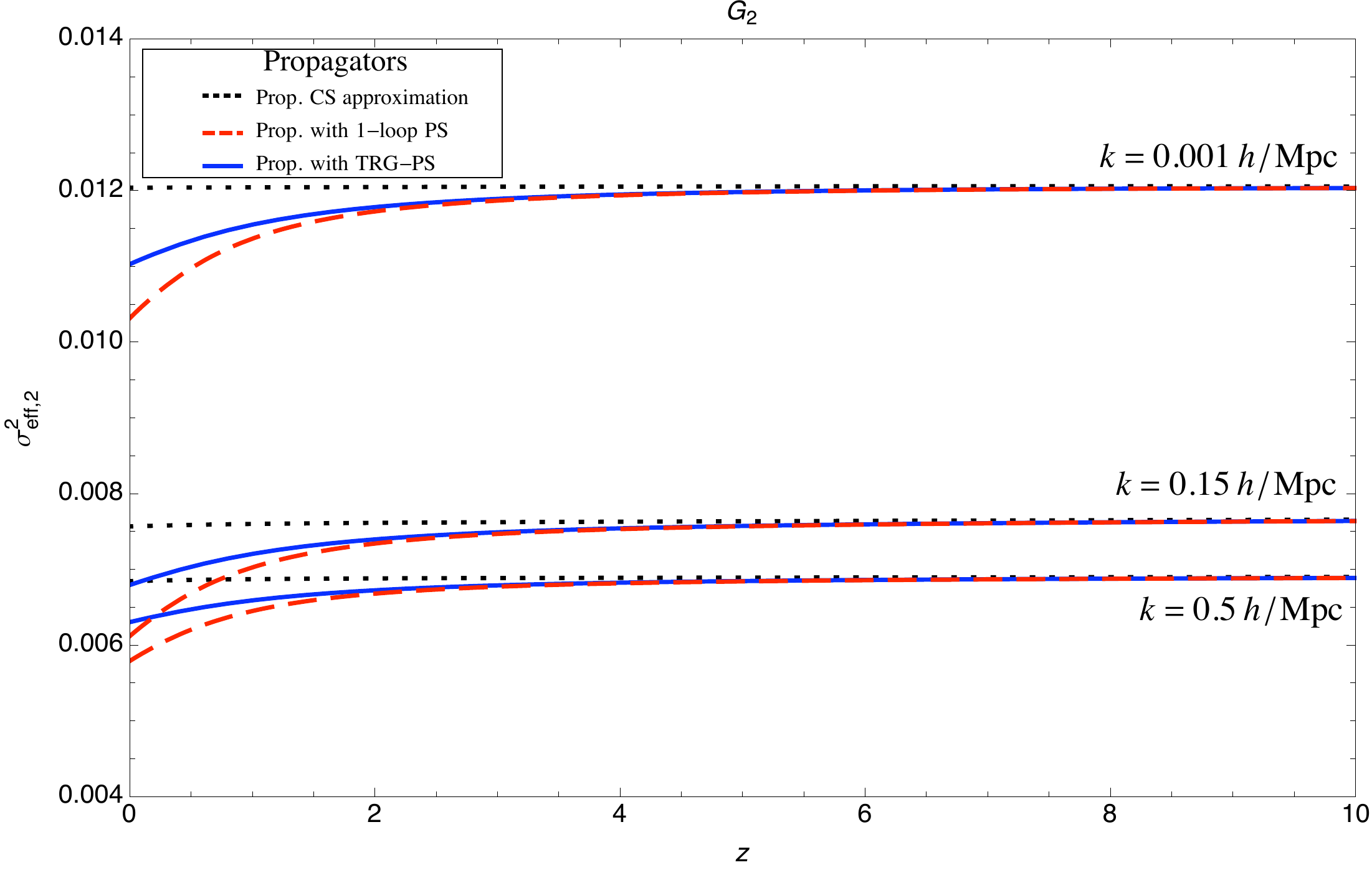}}
\caption{Same as Fig.~\ref{SIGMAG1Z} for  $\sigma_{eff,2}^2(z)$}
\label{SIGMAG2Z}
\end{figure}

Finally, in figs. \ref{SIGMAG1Z} and \ref{SIGMAG2Z} we plot $\sigma^{2}_{eff,a}$ as a function of redshift for three fixed values of the momentum:  $k=0.5,\, 0.15,\, 0.001\, \mathrm{h/Mpc}$. 

Our results show that, for $z\alt 2$, the subleading effects neglected in the Crocce-Scoccimarro approximation start to play a relevant role. For the density propagator $G_1$ they are larger that $1 \%$ for $k\ge 0.10 \, \mathrm{h/Mpc}$ at $z=0$ and for $k\ge 0.25 \,\mathrm{h/Mpc}$ at $z=1$. For the velocity propagator $G_2$ the effect is stronger. It is larger than $1 \%$ for $k\ge 0.07 \, \mathrm{h/Mpc}$ at $z=0$ and for $k\ge 0.16 \,\mathrm{h/Mpc}$ at $z=1$. These effects should clearly be taken into account in a computation aiming to reproduce the BAO power spectrum at the percent level. At $k=0.2 \, \mathrm{h/Mpc}$, that is, well inside the BAO range of scales, and at $z=0$, the deviation from the Crocce-Scoccimarro resummation is $4.0\,\%$ for $G_1$ and $7.8\,\%$ for $G_2$. 

\begin{figure}
\centerline{\includegraphics[width = 15cm,keepaspectratio=true]{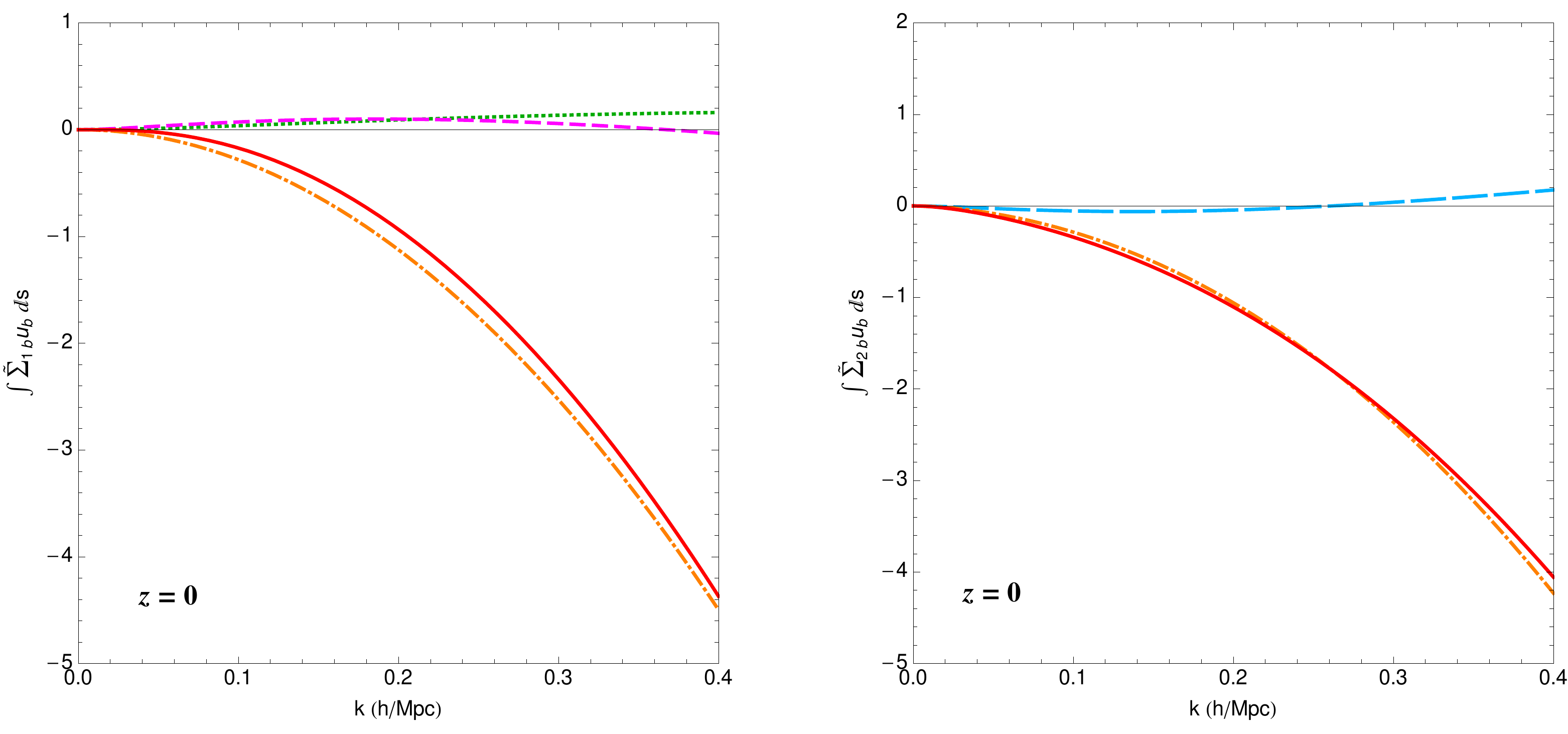}}
\caption{Contributions at $z=0$ to non linear term of Eq.~\re{TRGREN} for the density propagator (left panel) and velocity one (right panel). They are obtained by using 1-loop approximation for the non linear PS. 
The solid continuos lines give us the whole $\int \tilde{\Sigma}_{{\bf a}c}\,u_c$ value, while the dotted green, the short-dashed magenta, the long-dashed cyan and the dash-dotted orange ones represent the $P^\mathrm{nl}_{11}$, $P^\mathrm{nl}_{12}$,  $P^\mathrm{nl}_{21}$ and $P^\mathrm{nl}_{22}$ contributions respectively. }
\label{SigmaTildePS1loop}
\end{figure}

\begin{figure}
\centerline{\includegraphics[width = 15cm,keepaspectratio=true]{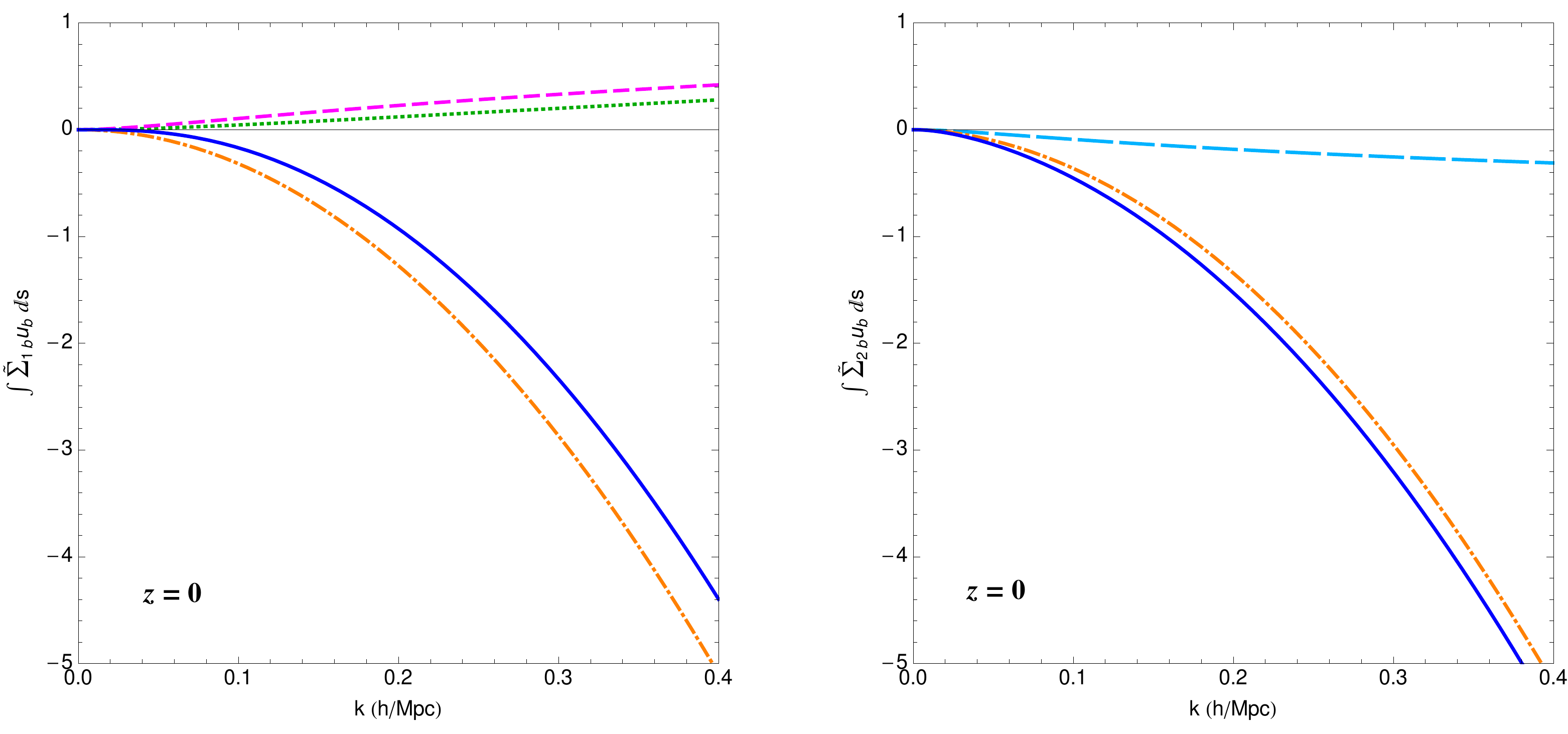}}
\caption{Same as Fig.~\ref{SigmaTildePS1loop} where now we consider the TRG approximation for the non linear PS.}
\label{SigmaTildePSTRG}
\end{figure}

The sign of the correction is the same in both approximations considered for the non-linear PS ({\it i.e.} 1-loop and TRG): the propagators are less damped than the Crocce-Scoccimarro one. This can be easily understood analytically by looking at the difference between Eq.~(\ref{resuL}) and Eq.~(\ref{resutNL}). In the former, the large-$k$ damping is modulated, via $\sigma^2$, by the linear PS, $P^0$, while, in the latter, it is modulated by the velocity-velocity component of the non-linear PS, {\it i.e.} by $P^\mathrm{nl}_{22}$, see Eq.~(\ref{sigt}). Now, unlike the non-linear density-density PS, $P^\mathrm{nl}_{11}$, which is enhanced w.r.t. the linear one, $P^\mathrm{nl}_{22}$ receives negative corrections, and is therefore smaller than $P^0$ at intermediate and large $k$'s (see, for instance, \cite{Pietroni:2008jx, Hiramatsu:2009ki}). Moreover in Eq.~(\ref{resutNL}) the unequal-time cross-correlator appears, which is further suppressed w.r.t. the equal-time one. As a consequence, we get $\sigma^2_\mathrm{nl}<\sigma^2$ and therefore a smaller damping at large $k$. On the other hand, for small and intermediate momentum values, the non linear part of Eq.~\re{TRGREN} depends also on $P^\mathrm{nl}_{11}$,  $P^\mathrm{nl}_{12}$ for the density propagator and on $P^\mathrm{nl}_{21}$ for the velocity propagator. In figs.~\ref{SigmaTildePS1loop} and \ref{SigmaTildePSTRG} we plot the contributions to $\int\, ds\, \tilde{\Sigma}_{{\bf a}c}\,u_c$ evaluated at redshift zero for 1-loop and TRG power spectra respectively. The solid continuos lines gives the whole $\int\, ds\, \tilde{\Sigma}_{{\bf a}c}\,u_c$ value, while the dotted green, the short-dashed magenta, the long-dashed cyan and the dash-dotted orange ones represent the $P^\mathrm{nl}_{11}$, $P^\mathrm{nl}_{12}$,  $P^\mathrm{nl}_{21}$ and $P^\mathrm{nl}_{22}$ contributions respectively. Notice that, also in this range of scales, the velocity-velocity component gives the dominant contribution to Eq.~\re{TRGREN}. In particular the non linear term that depends on the density-density component of the PS is positive but subdominant. Moreover, comparing the two figs.~\ref{SigmaTildePS1loop} and \ref{SigmaTildePSTRG} it is clear that the unphysical behavior of the 1-loop PS approximation discussed above translates into an underestimation of the non linear contributions to  Eq.~\re{TRGREN}.

\section{Discussion and conclusions}
\label{CONCL}
In this paper we have extended the computation of the non-linear propagator pioneered in CS, which was based on the resummation of the chain-diagrams at all orders in PT. 
We have taken into account new contributions, obtained by replacing the linear PS appearing in the chain-diagrams by the non-linear PS. We have proved the remarkable property (see \ref{fattorizzazione}), that this wider class of renormalized chain-diagrams can be exactly resummed in the large $k$ limit. In the same spirit of CS, we required that PT is recovered in the $k \to 0$ limit, which implies taking into account diagrams not belonging to the renormalized-chain diagrams class. 

The resummation of this extended class of diagrams is greatly simplified by the use of the time evolution equation for the full propagator, Eq.~(\ref{TRGREN}). Instead of dealing with a complex diagrammatic analysis, the task is reduced to the solution of a differential equation for the full propagator. The crucial element in this equation is the self-energy $\tilde{\Sigma}_{ab}$. Approximating it with the 1-loop self-energy gives the CS result. On the other hand, the renormalized chain-diagrams resummation is achieved by considering a $\tilde{\Sigma}_{ab}$ which is still formally 1-loop, but with the linear PS replaced by the non-linear one, see Fig.~\ref{SIGMA1}.
We have tested two different approximations for the non-linear PS: the 1-loop approximation, and the result of the TRG evolution \cite{Pietroni:2008jx}. The former gives results that depend sensibly on the UV cutoff in the loop integral. This is to be expected since, at low redshift, the 1-loop PS becomes unreliable, and even negative, al large momenta. On the other hand, the PS from the TRG evolution does not suffer from UV problems. Moreover, as discussed in detail in \cite{Pietroni:2008jx}, the solution of the TRG equations for the PS is formally a 1-loop expression, in which the linear PS' are replaced by non-linear ones, and is therefore fully consistent with the spirit of our treatment for the propagator in this paper.

The numerical results show that the new effects are quite relevant in the BAO scales, where they  are in the few percent range at z=0. They should therefore be taken into account in computations of the PS in the BAO range based on the use of renormalized propagators, such as RPT.
Indeed, in \cite{Crocce:2007dt}, the effect of next-to leading order corrections to the propagator was advocated in order to reconcile the RPT results on the PS with N-Body simulations. The authors correctly identified the effect of these corrections with a renormalization of the linear PS which, through the quantity  $\sigma^2$, modulates the Gaussian decay of the propagator at large momentum. However, instead of performing an explicit computation as the one presented in this work, they implemented an ad hoc procedure, by replacing the linear PS with the non-linear one as obtained in the halo model \cite{Scherrer1991, Cooray:2002dia}. This is inconsistent, since, by doing so, one is using the density PS, whereas the large-$k$ limit resummation involves the velocity PS, see Eq.~(\ref{largenl}). As a result, the procedure illustrated in \cite{Crocce:2007dt} leads to a wrong prediction on the sign of the corrections to the CS propagator induced by the subleading corrections. 

We stress that our conclusion that the effect of these corrections is to enhance the propagator w.r.t. the CS result is by no means based on a particular approximation of the non-linear PS. Different choices for the latter may give different results on the {\it size} of these corrections, but the sign is only determined by the assumption that the non-linear velocity PS is smaller than the linear one, which is verified in all consistent approximations (see, for instance, \cite{Pietroni:2008jx,Hiramatsu:2009ki})

A careful reconsideration of the comparison with N-body simulation is therefore needed, both for the density and for the velocity propagators. Should the discrepancy mentioned in  \cite{Pietroni:2008jx,Crocce:2007dt} persist, it would imply that other effects should be taken into account. One possibility would be to include diagrams not belonging to the renormalized chain class. At 2-loop order, it would mean to include diagrams VII, VIII, and IX in Fig.~\ref{FIGSIGMA2LOOP} also at large $k$. Notice that these contributions do not exponentiate in the large-$k$ limit or, equivalently, they break the factorization property of Eq.~(\ref{facto}) and, consequently, they give a propagator which deviates from the gaussian decay form found by CS at large $k$.  Another possible reason for the discrepancy could be the effect of small scale non-linearities, which translates in a non-vanishing velocity dispersion at intermediate scales see, for instance, \cite{Baumann:2010tm}. The inclusion of this effect in the computation of the (resummed) propagator will be analyzed elsewhere.

The approach followed in this paper is based on the use of the {\it exact}  evolution equation (\ref{FORMALTRGGEQ}).  
On the same spirit, an evolution equation for the PS can be written, and its results can be compared with alternative approaches presented in the recent literature \cite{Pietroni:2008jx,Taruya:2009ir,Hiramatsu:2009ki}. This will be the subject of a forthcoming publication.

\newpage

\appendix

\section{Factorization at large momentum} 
\label{fattorizzazione}
In this appendix, we will show that, at large momentum, Eq.~(\ref{facto}) holds, that is
\beqra
 && \sum_{j=0}^{n-1} \int_{\eta_b}^{\eta_a} d s\, \Sigma_{ac}^{(n-j)}( k;\,\eta_a\,,s)\,G_{cb}^{(j)}(k;\,s,\eta_b) \nonumber\\
 && \qquad \stackrel{\mathrm{large}\;k}{\longrightarrow} \;G_{{\bf a}b}^{(n-1)}(k;\,\eta_a,\eta_b) 
\int_{\eta_b}^{\eta_a} d s\, \Sigma_{{\bf a}c}^{(1)}( k;\,\eta_a\,,s)\,u_c\,.
\label{factoapp}
\eeqra
The first point to notice is that, as shown by CS in \cite{Crocce:2005xz}, the leading contributions at large $k$ and at a fixed loop order $n$ goes as $k^{2n}$, and is given by the chain-diagrams of Fig.~\ref{FIGCHAIN}. These diagrams are such that the $2n$ propagators are lined up in a single chain carrying the momentum $k$, and any of the $n$ power spectra, carrying a lower momentum $q_i$, is connected to the propagator chain by both its legs. Each of the $2 n$ vertices now contributes a factor
\beqra
 	   u_c\gamma_{acb}(\bk\,,-\bq\,,\bq-\bk)\stackrel{\mathrm{large}\,k}{\longrightarrow}\, \frac{1}{2}\frac{\bk\cdot\bq}{q^2}\, \delta_{ab}\,,
	   \label{lk} 
 \eeqra
where the $u_c$ comes from the PS (see Eq.~(\ref{s1l}) below), giving the above mentioned $O(k^{2n})$ behavior. Since the linear propagators are momentum-independent, the integrals over the loop momenta decouple one another, each one giving a contribution proportional to the 1-loop `self-energy',
\beqra
&&\Sigma^{(1)}_{a_i b_i}(k;\,s_{a_i},s_{b_i})=\, 4\,e^{s_{a_i}+s_{b_i}}  \,\int d^3q_i P(q_i) u_{c_i} u_{ e_i} \times \nonumber\\
&&  \gamma_{a_i c_i d_i}(\bk, \bq_i, -\bk-\bq_i) g_{d_i h_i}(s_{a_i}-s_{b_i})\gamma_{h_i e_i b_i}( \bk+\bq_i, -\bq_i, -\bk)\nonumber\\
&&\stackrel{\mathrm{large}\;k}{\longrightarrow} - k^2 \sigma^2 \,e^{s_{a_i}+s_{b_i}}   \,g_{a_i b_i}(s_{a_i}-s_{b_i})\,,
\label{s1l}
\eeqra
where 
\beq
\sigma^2\equiv \frac{1}{3} \int d^3 q \frac{P(q)}{q^2}\,,
\eeq
and we have used Eq.~(\ref{lk}).

Thanks to the composition property of the linear propagators 
\beq
g_{ac}(\eta_a-\eta_c)g_{cb}(\eta_c-\eta_b)=g_{ab}(\eta_a-\eta_b)\,,
\eeq
the chain of linear propagators emerging in the large $k$ limit of (\ref{s1l}) combine into a single one, $g_{ab}(\eta_a-\eta_b)$, independent of the intermediate times.

\begin{figure}
\centerline{\includegraphics[width = 15cm,keepaspectratio=true]{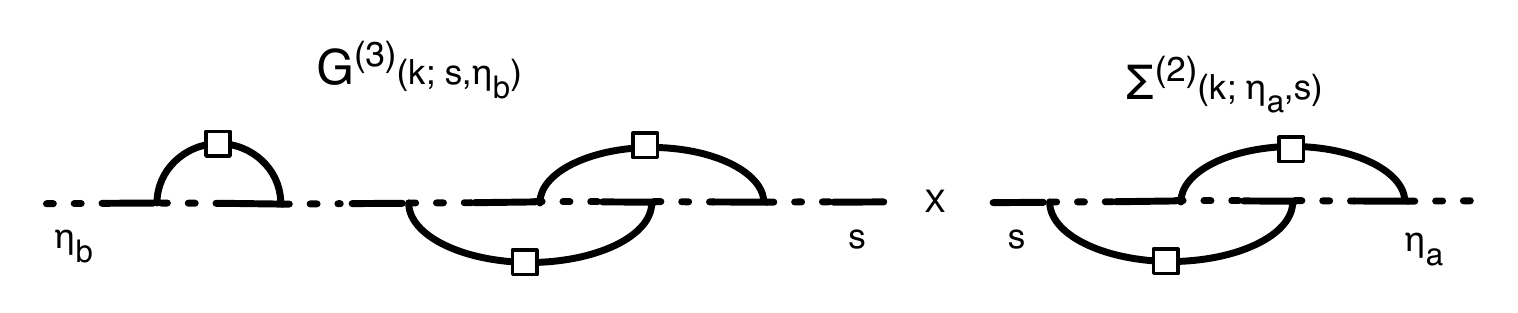}}
\caption{ A $5$-loop contribution to the sum of Eq. (\ref{factoapp}).}
\label{FIGAPPENDIX}
\end{figure}

In order to discuss the time integrals, we consider a generic $n$-loop contribution to the sum (\ref{factoapp}) (see Fig.~\ref{FIGAPPENDIX} for a $5$-loop example). The `self-energy' is a $n-j$ loop quantity and the propagator a $j$-loop one. Fixing the intermediate time $s$, the `self-energy' diagram has thus $2(n-j)-2$ intermediate times, while the propagator has $2j$.
The `self-energy' time integrals give
\beqra
&&\int_s^{\eta_a}dt_1\int_s^{t_1}dt_2\cdots \int_s^{t_{2(n-j)-3)}}dt_{2(n-j)-2} \;e^{\eta_a+\sum_{i=1}^{2(n-j)-2} t_i}\nonumber\\
&&= \frac{e^{\eta_a}(e^{\eta_a}-e^s)^{2(n-j)-2}}{(2n-2j-2)!}\,.
\label{se}
\eeqra
On the other hand, the propagator time integrals are
\beqra
 \int_{\eta_b}^s d\tau_1\int_{\eta_b}^{\tau_1} d\tau_2 \cdots \int_{\eta_b}^{\tau_{2j-1}} d\tau_{2j} \;e^{\sum_{k=1}^{2j} \tau_k} = \frac{(e^s-e^{\eta_b})^{2j}}{(2j)!}\,.
 \label{pr}
 \eeqra
 Multiplying (\ref{se}) by (\ref{pr}) and by the remainig $e^s$ time factor, and then integrating over $s$ from $\eta_b$ to $\eta_a$, as in the LHS of (\ref{factoapp}) one gets the time coefficient
 \beq
 \frac{e^{\eta_a}(e^{\eta_a}-e^{\eta_b})^{2n-1}}{(2n-1)!}\,,
 \label{tcof}
 \eeq
 which is independent of $j$, {\it i.e.} is the same for any term in the sum in Eq.~(\ref{factoapp}), and depends only on the total loop order, $n$. The sum over $j$ ensures that the $n$ power spectra are attached to the propagator chain in all possible ways, {\it i.e.} that all the pairings between the $2n$ vertices are taken into account. There are $(2n-1)!!$ such pairings,  so, using Eqs.~(\ref{s1l}) and (\ref{tcof}), we obtain the LHS of (\ref{factoapp}) in the large momentum limit,
 \beqra
&& \frac{e^{\eta_a}(e^{\eta_a}-e^{\eta_b})^{2n-1}}{(2n-1)!} (2n-1)!! (-k^2 \sigma^2)^{n} \nonumber\\
&&= \frac{1}{(n-1)!} \left[-k^2 \sigma^2 \frac{(e^{\eta_a}-e^{\eta_b})^2}{2}\right]^{n-1}\;(-k^2 \sigma^2) e^{\eta_a}(e^{\eta_a}-e^{\eta_b})  \nonumber\\
&&= \;\frac{1}{(n-1)!} \left[-k^2 \sigma^2 \frac{(e^{\eta_a}-e^{\eta_b})^2}{2}\right]^{n-1} \;\int_{\eta_b}^{\eta_a} ds \;\Sigma_{ac}^{(1)}(k;\,\eta_a,s) u_c \,.
\label{almost}
\eeqra
The contribution to the propagator from chain-diagrams at $(n-1)$-loop order can be computed straightforwardly, by using the same properties considered above  \cite{Crocce:2005xz}. The integration over the $2(n-1)$  intermediate times gives
\beq
\int_{\eta_b}^{\eta_a} dt_1\int_{\eta_b}^{t_1}dt_2\cdots\int_{\eta_b}^{t_{2n-3}}dt_{2n-2}e^{\sum_{i=1}^{2n-2}t_i} =  \frac{(e^{\eta_a}-e^{\eta_b})^{2n-2}}{(2n-2)!}\,.
\eeq
Since there are $(2n-3)!!$ chain-diagrams at $n-1$ order, the propagator in the large $k$ limit reads
\beq
G_{{\bf a}b}^{(n-1)}(k;\,\eta_a,\eta_b) \stackrel{\mathrm{large}\;k}{\longrightarrow}   \;\frac{1}{(n-1)!} \left[-k^2 \sigma^2 \frac{(e^{\eta_a}-e^{\eta_b})^2}{2}\right]^{n-1} \, g_{{\bf a}b}(\eta_a-\eta_b),
\label{propL}
\eeq
which, comparing with the last line of (\ref{almost}), proves Eq.~(\ref{factoapp}).

The factorization property, Eq.~(\ref{factoapp}), is a general property of chain-diagrams,  which does not rely on the fact that the PS considered above is the linear one: $P_{ab}(q; s_a,s_b) = P^0(q) u_a u_b$. Indeed, in the remaining part of this Appendix, we will show that the factorization still holds if one considers a generic form for the PS, {\it i.e.} 
\beq
P_{ab}(q; s_a,s_b) = P_{ba}(q; s_b,s_a) \,,
\eeq
and therefore it holds if one renormalizes the linear PS by including non-linearities in different consistent approximations, such as, perturbation theory, TRG, and so on.

The starting point is to realize that the LHS of  Eq.~(\ref{factoapp})  is obtained by pairing in all possible ways the $2n$ vertices -- including the extremal one at time $\eta_a$ -- connected by the chain of $2n$ propagators. Moreover, the large momentum property of the vertex, Eq.~(\ref{lk}), still holds of it is contracted by a generic vector $A_a$, giving
\beqra
 	   A_c\gamma_{acb}(\bk\,,-\bq\,,\bq-\bk)\stackrel{\mathrm{large}\,k}{\longrightarrow}\, A_2\,  \frac{1}{2}\frac{\bk\cdot\bq}{q^2}\,\delta_{ab}\,.
	   \label{lkA} 
 \eeqra
 Therefore, in the large $k$ limit, we have \footnote{Now the upper indices count the number of non-linear PS in a given contribution to $\Sigma$ and $G$.}
 \beqra
 && \sum_{j=0}^{n-1} \int_{\eta_b}^{\eta_a} d s\, \Sigma_{ac}^{(n-j)}( k;\,\eta_a\,,s)\,G_{cb}^{(j)}(k;\,s,\eta_b) \nonumber\\
 && \stackrel{\mathrm{large}\;k}{\longrightarrow} 
\left(\frac{-k^2}{3}\right)^n  \int_{\eta_b}^{\eta_a} ds_1\int_{\eta_b}^{s_1} ds_2\cdots\int_{\eta_b}^{s_{2n-2}} ds_{2n-1} \,\left(\int \Pi_{i=1}^n d^3 q_i \right) \times \nonumber \\
&&e^{\eta_a+\sum_{i=1}^{2n-1} s_i}   \sum_{2n\, \mathrm{pairings}} \frac{P_{22}(q_1;\eta_a, s_{a_1})}{q_1^2}\cdots
\frac{P_{22}(q_n;s_{a_{2n-2}}, s_{a_{2n-1}})}{q_n^2}\,.\nonumber\\
&&
\label{A11}
 \eeqra 
 Notice that the time-integrand is, by construction, invariant under the exchange of any of the $2n-1$ variables, $s_i \leftrightarrow s_j$, therefore we will use the property
 \beqra
&&  \int_{\eta_b}^{\eta_a} ds_1\int_{\eta_b}^{s_1} ds_2\cdots\int_{\eta_b}^{s_{N-1}} ds_{N} \,{\cal F}[s_1, \cdots,s_N] = \nonumber\\
&&\qquad\qquad\qquad\qquad \frac{1}{N!} 
    \int_{\eta_b}^{\eta_a} ds_1\int_{\eta_b}^{\eta_a}ds_2\cdots\int_{\eta_b}^{\eta_a} ds_{N}\,{\cal F}[s_1, \cdots,s_N]\,,
  \eeqra
  where the function ${\cal F}[s_1, \cdots,s_N]$ is totally symmetric.
 Eq.~(\ref{A11})  can then be rewritten as
 \beqra
&&\left(\frac{-k^2}{3}\right)^n \frac{1}{(2n-1)!} \int_{\eta_b}^{\eta_a} ds_1\int_{\eta_b}^{\eta_a} ds_2\cdots\int_{\eta_b}^{\eta_a} ds_{2n-1} \,\left(\int \Pi_{i=1}^n d^3 q_i \right) \times \nonumber \\
&&e^{\eta_a+\sum_{i=1}^{2n-1} s_i}   \sum_{2n\, \mathrm{pairings}} \frac{P_{22}(q_1;\eta_a, s_{a_1})}{q_1^2}\cdots
\frac{P_{22}(q_n;s_{a_{2n-2}}, s_{a_{2n-1}})}{q_n^2} =\nonumber \\
&& \left(\frac{-k^2}{3}\right) \int_{\eta_b}^{\eta_a}ds \, e^{\eta_a+s} \int d^3 q \frac{P_{22}(q;\eta_a, s)}{q^2} \times \nonumber\\
&&\left[\left(\frac{-k^2}{3}\right)^{n-1} \frac{1}{(2n-2)!} \int_{\eta_b}^{\eta_a} ds_1\int_{\eta_b}^{\eta_a} ds_2\cdots\int_{\eta_b}^{\eta_a} ds_{2n-2} \,\left(\int \Pi_{i=1}^{n-1} d^3 q_i \right) \times \right.\nonumber \\
&&\left. e^{\sum_{i=1}^{2n-2} s_i} \sum_{2n-2\, \mathrm{pairings}} \frac{P_{22}(q_1;s_{a_1},s_{a_2})}{q_1^2}\cdots
\frac{P_{22}(q_{n-1};s_{a_{2n-3}}, s_{a_{2n-2}})}{q_{n-1}^2}\right]\,,\nonumber\\
&&
 \eeqra 
 which, multiplied by $g_{ab}(\eta_a-\eta_b)$, gives the RHS of (\ref{factoapp}) (see also (\ref{s1l}) and (\ref{propL})).

\section{Time-Remornalization Group equations for the non linear PS} 
\label{PHIPHIAPP}
In this Appendix we briefly review the Time-Remornalization Group approach (TRG) introduced in \cite{Pietroni:2008jx} to compute the non linear PS. Applying the equation of motion in Eq.~\re{compact} to the (equal-time) PS, the bispectrum,
\beq
\langle \vp_a(\bk,\eta)\vp_b(\bp,\eta) \vp_c(\bq,\eta) \rangle = \delta_D(\bk+\bp+\bq)\,B_{abc}(\bk,\bp,\bq;\eta)\,,
\eeq
and to the higher order correlators, one gets an infinite system of coupled differential equations. Truncating the hierarchy by setting the trispectum ({\it i.e.} the connected four-point function) to zero, one is left with the closed system
\beqra
&&\ds  \partial_\eta\,P_{ab}({\bf k}\,;\eta) = - \Omega_{ac}P_{cb}({\bf k}\,;\eta)  - \Omega_{bc} P_{ac}({\bf k}\,;\eta) \nonumber\\
&&\qquad\qquad\quad\quad+e^\eta \int d^3 q\, \left[ \gamma_{acd}({\bf k},\,{\bf -q},\,{\bf q-k})\,B_{bcd}({\bf k},\,{\bf -q},\,{\bf q-k};\,\eta)\right.\nonumber\\
&&\qquad\qquad\qquad\qquad\qquad\left. + B_{acd}({\bf k},\,{\bf -q},\,{\bf q-k};\,\eta)\,\gamma_{bcd}({\bf k},\,{\bf -q},\,{\bf q-k})\right]\,,\nonumber\\
&&\nonumber\\
&&\ds  \partial_\eta\,B_{abc}({\bf k},\,{\bf -q},\,{\bf q-k};\,\eta) =  - \Omega_{ad} B_{dbc}({\bf k},\,{\bf -q},\,{\bf q-k};\,\eta)\nonumber\\
&&\qquad\qquad\qquad\qquad\qquad\quad- \Omega_{bd} B_{adc}({\bf k},\,{\bf -q},\,{\bf q-k};\,\eta)\nonumber\\
&&\qquad\qquad\qquad\qquad\qquad\quad - \Omega_{cd}B_{abd}({\bf k},\,{\bf -q},\,{\bf q-k};\,\eta)\nonumber\\
&&\qquad\qquad\qquad\qquad + 2 e^\eta \left[ \gamma_{ade}({\bf k},\,{\bf -q},\,{\bf q-k}) P_{db}({\bf q}\,;\eta)P_{ec}({\bf k-q}\,;\eta)\right.\nonumber\\
&&\qquad\qquad\qquad\qquad\quad +\gamma_{bde}({\bf -q},\,{\bf q-k},\,{\bf k}) P_{dc}({\bf k-q}\,;\eta)P_{ea}({\bf k}\;;\eta)\nonumber\\
&&\qquad\qquad\qquad\qquad\quad +\left. \gamma_{cde}({\bf q-k},\,{\bf k},\,{\bf -q}) P_{da}({\bf k}\,;\eta)P_{eb}({\bf q}\,;\eta)\right]\,.
\label{systPS}
\eeqra
The formal solution of the system (\ref{systPS}) is given by
\beqra
&& P_{ab}({\bf k}\,;\eta) = g_{ac}(\eta,0)  \, g_{bd}(\eta,0)  P_{cd}({\bf k}\,;\eta=0) \nonumber\\
&&  
\qquad\qquad\quad+\int_0^\eta d\eta^\prime e^{\eta^\prime} \int d^3 q \,g_{ae}(\eta,\eta^\prime) g_{bf}(\eta,\eta^\prime) \nonumber\\
&&\qquad\qquad\qquad\quad\ \times\left[  \gamma_{ecd}({\bf k},\,{\bf -q},\,{\bf q-k})\,B_{fcd}({\bf k},\,{\bf -q},\,{\bf q-k};\,\eta^\prime)\right.\nonumber\\
&&\qquad\qquad\qquad\qquad \quad+
\left. \gamma_{fcd}({\bf k},\,{\bf -q},\,{\bf q-k})\,B_{ecd}({\bf k},\,{\bf -q},\,{\bf q-k};\,\eta^\prime)\right]\,,\nonumber\\
&&\nonumber\\
&&B_{abc}({\bf k},\,{\bf -q},\,{\bf q-k};\,\eta)=\nonumber\\
&&
 \qquad g_{ad}(\eta,0)g_{be}(\eta,0) g_{cf}(\eta,0)B_{def}({\bf k},\,{\bf -q},\,{\bf q-k};\,\eta=0)\nonumber\\
&&
 \qquad\qquad\qquad +2 \int_0^\eta d\eta^\prime e^{\eta^\prime} \,g_{ad}(\eta,\eta^\prime) g_{be}(\eta,\eta^\prime) g_{cf}(\eta,\eta^\prime)\nonumber\\
&&
\qquad \qquad\qquad\qquad \times\left[ \gamma_{dgh}({\bf k},\,{\bf -q},\,{\bf q-k})P_{eg}({\bf q}\,;\eta^\prime)P_{fh}({\bf q-k}\,;\eta^\prime)\right.\nonumber\\
&& \quad\qquad\qquad\qquad\qquad + \gamma_{egh}({\bf -q},\,{\bf q-k},\,{\bf k})P_{fg}({\bf q-k}\,;\eta^\prime)P_{dh}({\bf k}\,;\eta^\prime)\nonumber\\
&& \quad\qquad\qquad\qquad\qquad  \left.
+ \gamma_{fgh}({\bf q-k},\,{\bf k},\,{\bf -q})P_{dg}({\bf k}\,;\eta^\prime)P_{eh}({\bf q}\,;\eta^\prime)
\right]\,,
\label{formalsolPS}
\eeqra
which shows that the non-linear PS in this approach is given by a formally 1-loop expression in which non-linear PS' replace the linear ones. In this respect, the TRG approach is fully consistent with the computation of the propagator presented in this paper, which is based, as well, on improving over the CS approximation by replacing linear PS' with non-linear ones.

\section*{Acknowledgments}
We thank G. Ballesteros for the useful discussions.

\section*{References}
\bibliographystyle{JHEP}
\bibliography{PropBibl}
\end{document}